\documentclass[12pt]{article}

\usepackage{newtxtext,newtxmath}
\usepackage{graphicx}
\usepackage[letterpaper,margin=1in]{geometry}

\renewenvironment{abstract}
	{\quotation}
	{\endquotation}

\date{}

\makeatletter
\renewcommand{\fnum@figure}{\textbf{Figure \thefigure}}
\renewcommand{\fnum@table}{\textbf{Table \thetable}}
\makeatother

\usepackage{scicite}

\usepackage{url}
\usepackage{color}
\usepackage{braket}
\usepackage{dsfont}
\usepackage{hyperref}
\renewcommand{\Re}{\textrm{Re}}

\def\scititle{In-situ Nonlinear Adjoint Propagation for Multi-Path Wave Control	
}
\title{\bfseries \boldmath \scititle}

\author{
	John Guillamon$^{1\dagger}$,
    William Tuxbury$^{1\dagger}$,
	Cheng-Zhen Wang$^{1\dagger}$,
	Owen Miller$^2$,\\
	Zin Lin$^3$,
	Tsampikos Kottos$^{1\ast}$\and
	\small$^{1}$Wave Transport in Complex Systems Lab, Department of Physics, Wesleyan University, Middletown, CT-06459, USA.\and
	\small$^{2}$Department of Applied Physics, yale University, New Haven, 06520, CT, USA.\and
	\small$^3$Department of Electrical and Computer Engineering, Virginia Tech, Blacksburg, 24060, VA, USA\and
	\small$^\ast$Corresponding author. Email:  tkottos@wesleyan.edu\and
	\small$^\dagger$These authors contributed equally to this work.
}

\begin{document}

\maketitle
\begin{abstract}\bfseries\boldmath
Complex multipath environments are usually avoided in wave-based information processing because repeated scattering creates many 
interfering propagation paths, obscuring controllability and generating extreme sensitivity to perturbations. The addition of nonlinear mechanisms 
fundamentally alters the wave-control landscape by breaking the superposition principle that underpins most wave-management strategies. Here, 
we show that these two apparent impediments -- multipath complexity and nonlinearity -- can instead be harnessed as key resources for physical 
optimization. We demonstrate an in-situ adjoint optimization protocol in a wave-chaotic platform incorporating a single localized nonlinear 
defect, in which the system itself performs both the forward and the adjoint propagations required for gradient evaluation. Recurrent multipath 
returns repeatedly expose the wave to the defect, producing from a minimal hardware a rich nonlinear input-output map with many pathway-mediated 
degrees of freedom. At the same time, a suitable adjoint excitation enables direct extraction of the sensitivities from measurements alone, 
without a digital twin or conventional numerical backpropagation. We experimentally validate the protocol on a minimal nonlinear multipath 
platform composed of incommensurate coaxial cables connected via T-junctions, one of which hosts a diode-loaded cavity. Our approach opens a 
route to adaptive wireless communications, imaging and analog intelligence in complex, partially unknown environments where conventional modeling 
is impractical.
\end{abstract}



\maketitle

\section*{Introduction}

Wave control lies at the heart of a wide range of physical phenomena and technological platforms, from electromagnetic propagation in photonic 
and microwave circuits to acoustic and mechanical vibrations in resonator arrays. Even in linear systems, multiple scattering gives rise to highly structured 
and intricate behaviors, including interference patterns, resonant enhancement, wave localization, and anomalous transport~\cite{LTW2009,
CMR2022,gigan2022}. These effects have been harnessed for many applications such as wave focusing~\cite{Fink2,Fink1,BYGYHC2022,
MGBYHYC2024,hsu2017correlation}, coherent perfect absorption \cite{WCGNSC2011,CKA2020,FSK2017,aluCPA}, imaging through disordered media~\cite{BGH2022,SLM1,cizmar2015,MGBYHYC2024,KSBS11}, signal routing~\cite{CMR2022,WGKDRGK26,jiang2024,RIS7,YDLWS25}, 
and analog computation~\cite{wetzstein2020,SMCGAE2014,ZSAF2021,Solli2015}. 

Despite these successes, linear scattering is fundamentally constrained by the superposition principle, which imposes strict limitations on the range 
of accessible wave phenomena. In particular, the linear response of a system prevents self-induced adaptation or nonlinear feedback, limits the controllability 
of multi-wave interactions, and constrains the diversity of achievable functionalities. As a result, while linear platforms continue to enable powerful 
and elegant physical effects, they are intrinsically incapable of realizing the full spectrum of complex, adaptive, or highly nonlinear behaviors observed 
in many natural and engineered systems. Not only are many physical systems inherently nonlinear, but nonlinear multiple scattering dramatically expands 
the landscape of accessible behaviors \cite{SM2003}, giving rise to even richer dynamical complexity and enabling more advanced functionalities across 
diverse applications, including noise squeezing and enhanced sensing~\cite{SFKK2023,Chen2024,SFWRKK2025}, complex directional energy 
transport~\cite{WGTKK24,Galiffi2026,GHRSD2024,WKKK23,HMWF18}, as well as universal computation and physical deep learning~\cite{wright2022,
shen2017deep,Fleury2025,Momeni2023,WSWSMTAM2023,Papp2021}. 

Yet, this very complexity poses a major challenge for accurate modeling and rapid \textit{in-silico} simulation and optimization of targeted functionalities. 
A central obstacle is the simulation–reality gap: in complex physical platforms, even weak discrepancies between the simulated model and the realized device 
can corrupt gradient evaluation and drive optimization toward solutions that do not transfer reliably to experiment. In most existing approaches, backscattering, 
parasitic reflections, calibration errors and other hardware-specific irregularities can plague the optimization process and are therefore regarded as 
imperfections to be minimized. 

An alternative to purely \textit{in-silico} approaches is the emerging idea of \textit{in-situ} nonlinear backpropagation~\cite{GBWL21,SGL25}, in which 
a \textit{reconfigurable} physical system is optimized directly in hardware and in real time, for on-demand functionalities -- entirely bypassing inaccurate 
or inefficient digital twins. Yet this program remains largely underdeveloped: existing demonstrations are confined to narrow feed-forward settings with highly specialized nonlinearities~\cite{GBWL21,SGL25},  
where scattering complexity itself is treated as a nuisance, rather than a computational resource. Similarly, the few in-situ optimization schemes that implement 
the well-known ``adjoint'' method -- leveraging one forward and one backward pass through a system to enable the \textit{simultaneous} measurement of 
all gradient sensitivities -- rely on time-reversal symmetry \cite{hughes2018,PSHPBWMMA2023,LM23} or (at least) reciprocity~\cite{DMW2025,GWLK2025}, 
which permit an exact mapping between the adjoint dynamics and the original ``forward'' physical system. Many practical nonlinear platforms, by contrast, 
inevitably break time-reversal symmetry (e.g., due to material absorption) or inherently violate reciprocity due to nonlinearities, thereby precluding 
in-situ optimization and raising the fundamental question of whether the in-situ optimization paradigm can be implemented to such physically important systems. 

Here, we address this fundamental challenge by considering a minimal platform consisting of a single nonlinear defect embedded in a multipath and multiresonant 
network of coaxial cables that supports wave-chaotic dynamics \cite{kottos2000}. A nonlinear defect in such a complex multipath environment constitutes 
a prototypical building block: the recurrent scattering events cause waves to encounter the defect repeatedly, producing an increasingly complex nonlinear 
input-output map with many pathway-mediated degrees of freedom from minimal hardware. We show that the adjoint solution may be physically evaluated by injecting 
a perturbative signal under sustained forward excitation. It turns out that the resulting perturbed field generally acquires undesired phase discrepancies, 
compared to the 
true adjoint solution, arising from the non-reciprocal nature of nonlinear dynamics. We overcome this obstacle by appropriately designing the phase of the 
adjoint source such that the field at the nonlinear defect remains invariant under phase 
conjugation. This invariance enables the adjoint dynamics of the nonlinear wave equation to be exactly emulated by the same hardware as the original 
physical system. The methodology is validated through proof-of-concept experimental demonstrations on a microwave network testbed, achieving real-time 
\textit{in-situ Nonlinear Adjoint Propagation} (iNAP) for the optimization of on-demand nonlinear functional modalities, including power splitters, 
nonlinear coherent perfect absorbers, and asymmetric transport. 

\begin{figure}[htbp]
\centering
\includegraphics[width=0.9\textwidth]{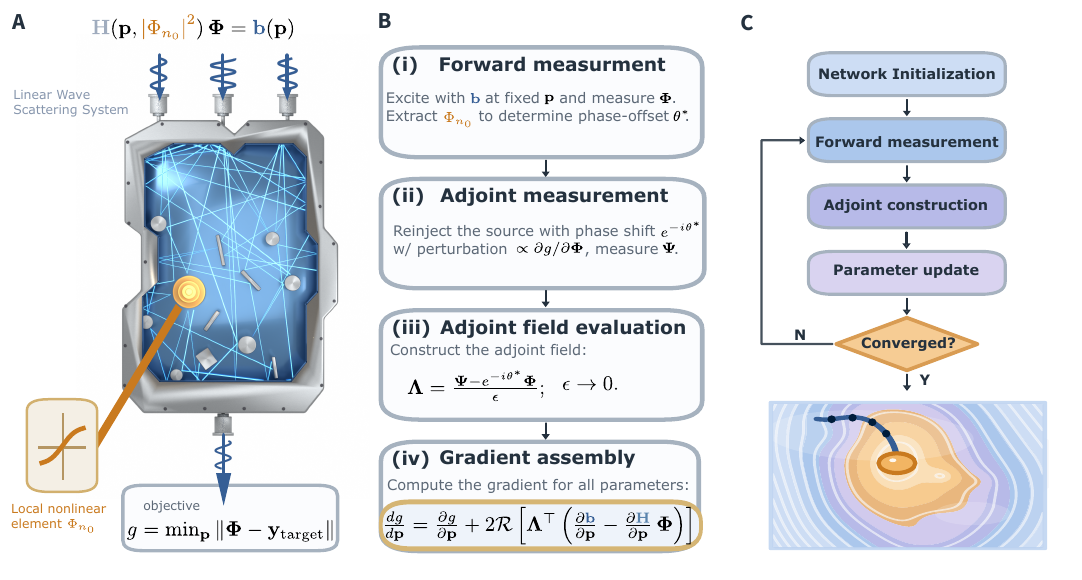}
\caption{\textbf{In-situ nonlinear adjoint protocol for gradient-based control of a wave-scattering system.}
\textbf{(A)} A physical wave-scattering network with tunable control parameters $\mathbf{p}$ and a local nonlinear element $\Phi_{n_0}$ is described by
$\mathbf{H}(|\Phi_{n_0}|^2,\mathbf{p})\,\mathbf{\Phi}=\mathbf{b}(\mathbf{p})$. The measured field response $\mathbf{\Phi}$ defines an objective $g$, 
illustrated here as the mismatch to a target output. \textbf{(B)} The iNAP protocol: (i) {\it Forward measurement}: Set the current control parameters 
$\mathbf{p}$, excite the physical system with $\mathbf{b}$, and measure the excited field ${\mathbf \Phi}$ at targeted positions, including $\Phi_{n_0}$ 
at the position of the nonlinear element. Use the measured nonlinear-site field $\Phi_{n_0}$ to determine the global phase correction factor $\theta^*$. 
(ii) {\it Adjoint measurement:} With the control parameters $\mathbf{p}$ fixed, inject a phase-engineered backward source $\Tilde{\mathbf{b}}$ by phase-
rotating the original forward excitation $\mathbf{b}$ by $\theta^*$ and adding a small perturbation with amplitude $\epsilon\propto \partial g/\partial
\mathbf{\Phi}$, and measure the excited field $\mathbf{\Psi}$. (iii) {\it Adjoint field evaluation}: Evaluate the adjoint field $\mathbf{\Lambda}$. (iv) 
{\it Gradient assembly}: Use the adjoint formula to assemble the gradient with respect to all control parameters simultaneously. \textbf{(C)} Iterative 
gradient-based optimization loop combining network initialization, forward measurement, adjoint construction and parameter update until convergence, 
illustrated as descent on the objective landscape.}\label{Fig1}
\end{figure}

\section*{Results}
\subsection*{Principles of In-Situ Nonlinear Adjoint Optimization}
We consider a lossy scattering setup with a single localized nonlinear defect and no microscopic time-reversal-symmetry-breaking fields such as 
magnetic bias, see Fig. \ref{Fig1}a. While the linear background is microscopically unbiased, the nonlinear defect can generate nonreciprocal scattering. The steady-state 
wave field $\mathbf{\Phi}(\mathbf{p})$, which depends on $N$ tunable optimization parameters $\mathbf{p}\equiv (p_1,p_2,\cdots,p_N)^T$, is 
determined by a nonlinear equation (e.g., a self-consistent Maxwell equation with intensity-dependent impedance profile) $\mathbf{F}
(\mathbf{p},\,\mathbf{\Phi},\,\mathbf{\Phi^*}) \;=\; \mathbf{H}(\mathbf{p},\, \mathbf{\Phi},\mathbf{\Phi^*})\,\,\mathbf{\Phi} - \mathbf{b}(\mathbf{p}) 
\;=\; 0$, where $\mathbf{b}$ is the driving source. Here $\mathbf{H}=\mathbf{H}_{\text L}+\mathbf{H}_{\text {NL}}$, where $\mathbf{H}_{\text L}$ is a linear operator 
describing the background environment and $\mathbf{H}_{\text {NL}}=f(|\Phi_{n_0}|^2) \mathbf {P}_{n_0}$ is a nonlinear operator that depends on the local 
field intensity $|\Phi_{n_0}|^2$ through a nonlinear defect acting at position $n_0$, and $(\mathbf{P}_{n_0})_{nm}=\delta_{n,n_0}\delta_{n,m}$ is 
a projection operator at $n_0$. Rather than assuming a particular nonlinear response, as in Ref. \cite{GBWL21,SGL25}, our protocol remains 
agnostic to the specific functional form of the nonlinearity, treating $f(|\Phi_{n_0}|^2)$ in full generality.

The optimization objective $g(\mathbf{\Phi}, \mathbf{\Phi^*},\mathbf{p})$ depends on $\mathbf{\Phi}$, $\mathbf{\Phi^*}$, and $\mathbf{p}$. The gradient 
sensitivities of $g$ with respect to $\mathbf{p}$ are:
\begin{align}
    \frac{d g}{d \mathbf{p}}
    \;=\;
    \frac{\partial g}{\partial \mathbf{p}}
    \;-\;
    \begin{bmatrix}
        \displaystyle \frac{\partial g}{\partial \mathbf{\Phi}},
        \displaystyle \frac{\partial g}{\partial \mathbf{\Phi}^*}
    \end{bmatrix}
    {\mathcal J}^{-1}
\begin{bmatrix}
\dfrac{\partial \mathbf{F}}{\partial \mathbf{p}} \\[6pt]
\dfrac{\partial \mathbf{F}^*}{\partial \mathbf{p}}
\end{bmatrix};\quad {\mathcal J}=\begin{bmatrix}
\dfrac{\partial \mathbf{F}}{\partial {\mathbf \Phi}} & \dfrac{\partial \mathbf{F}}{\partial {\mathbf \Phi}^*} \\[6pt]
\dfrac{\partial \mathbf{F}^*}{\partial {\mathbf \Phi}} & \dfrac{\partial \mathbf{F}^*}{\partial {\mathbf \Phi}^*}
\end{bmatrix},
\end{align}
where $\partial g/\partial\mathbf{ \Phi}$ denotes a row-vector of sensitivities at each field component and $\mathcal{J}$ is the Wirtinger Jacobian 
associated with the nonlinear equation $\mathbf{F}(\mathbf{p},\,\mathbf{\Phi},\,\mathbf{\Phi^*})=0$, evaluated at the steady-state solution $\mathbf
{\Phi}$. For the upper left submatrix we have ${\mathcal J}_{11} \equiv \dfrac{\partial \mathbf{F}}{\partial {\mathbf \Phi}}=\mathbf{H}_L+
\left[f({\cal I}_{n_0})+f'({\cal I}_{n_0}){\cal I}_{n_0}\right]\mathbf{P}_{n_0}$, where the prime indicates derivative with respect to ${\cal I}_{n_0}=|\Phi_{n_0}|^2$. 
Similarly the upper right block is $\mathcal {J}_{1,2}\equiv  \dfrac{\partial \mathbf{F}}{\partial {\mathbf \Phi}^*}=f'({\cal I}_{n_0})\Phi_{n_0}^2
\mathbf{P}_{n_0}$. The blocks in the lower row are $\mathcal{J}_{2,1}=\mathcal{J}_{1,2}^*, \mathcal{J}_{2,2}=\mathcal{J}_{1,1}^*$. In the 
weak-field limit, the response of the local defect is effectively linear and the Jacobian becomes block-diagonal, i.e., $\mathcal{J}_{11}=\mathbf{H}_L, 
\mathcal{J}_{12}=\mathcal{J}_{2,1}=0$ and $\mathcal{J}_{2,2}=\mathbf{H}_L^*$, where $\mathbf{H}_L=\mathbf{H}_L^T$ for microscopic 
time-reversible systems. At higher incident fields, activation of the local nonlinearity breaks this simple structure by introducing an explicit dependence 
on both the amplitudes and their complex conjugates. The defect then contributes not only to the diagonal block, but also to an anomalous off-diagonal 
term proportional to $f'({\cal I}_{n_0})\Phi_{n_0}^2$.

Finally, $\left(\dfrac{\partial \mathbf{F}}{\partial \mathbf{p}};\dfrac{\partial \mathbf{F}^*}{\partial \mathbf{p}}\right)^\top$ physically represents a 
collection of induced {\it small} excitations resulting from perturbing the ({\it linearized}) system via one parameter $p_i$ at a time for each $i=1,2,
\cdots,N$. Consequently, $\mathbf{U}=\mathcal{J}^{-1}\left(\dfrac{\partial \mathbf{F}}{\partial \mathbf{p}};\dfrac{\partial \mathbf{F}^*}{\partial \mathbf{p}}
\right)^\top$ denotes a collection of several wave fields in response to each of these perturbations (the $i$-th column, $\mathbf{U}_i$, corresponds 
to the wave field generated by perturbing the single $p_i$). However, finding the entire $\mathbf{U}$ becomes excessive, especially when the number 
of controllable parameters, $N$, is large.

An in-silico resolution to such inefficient and time-consuming approaches is provided by the adjoint method, which reformulates the optimization problem 
as an auxiliary backward-propagation problem for efficient gradient evaluation \cite{HMWF18}. To this end, we introduce an adjoint field $\mathbf{\Lambda}$ 
and its complex-conjugate $\mathbf{\Lambda}^*$, that satisfies a {\it linear} adjoint equation (even though the physical problem itself is nonlinear) 
$\Big(\frac{\partial \mathbf{F}}{\partial \mathbf{\Phi}}\Big)^\top \mathbf{\Lambda} \;+\; \Big(\frac{\partial \mathbf{F}^*}{\partial \mathbf{\Phi}}\Big)^\top 
\mathbf{\Lambda} ^* \;=\; -\, \left(\frac{\partial g}{\partial \Phi}\right)^\top$ and its complex conjugate. Such equations result from a linearization of 
the system equations about the operating point of the forward propagation. The linear adjoint equation can be written as:
\begin{align}
   \Big(\mathbf{H}_L+[f({\cal I}_{n_0})+{\cal I}_{n_0}f'({\cal I}_{n_0})]\mathbf{P}_{n_0}\Big) \mathbf{\Lambda} \;+\; f'({\cal I}_{n_0})^*(\Phi_{n_0}^*)^2 \mathbf{P}_{n_0}
   \mathbf{\Lambda} ^* \;=\; -\,\left(\frac{\partial g}{\partial \mathbf{\Phi}}\right)^\top,
\label{eq:nladjoint}
\end{align}
where we have used that $\mathbf{H}_L^\top=\mathbf{H}_L$ and $\mathbf{P}_{n_0}^\top=\mathbf{P}_{n_0}$.
Once $\mathbf{\Lambda}$ is obtained, the gradient of the objective with respect to the parameters can be evaluated as:
\begin{align}
    \frac{d g}{d \mathbf{p}} \;=\; \frac{\partial g}{\partial \mathbf{p}} \;+\; 2\,\Re\!\Big\{\mathbf{\Lambda}^\top \Big(\frac{\partial \mathbf{F}}
    {\partial \mathbf{p}}\Big)\Big\},
\label{eq:nlgrad}
\end{align}
where we have implicitly assumed that both $g$ and $\mathbf{p}$ are real. 
This formulation significantly reduces computational demands, as all sensitivities can be obtained through a single additional {\it linear} field 
propagation, instead of computing the entire collection $\mathbf{U}$ of $N$ wave fields -- thus making large-scale, gradient-based optimization 
possible. Importantly, $\Big(\frac{\partial \mathbf{F}}{\partial \mathbf{p}}\Big)$ is typically a very sparse tensor since the effect of each parameter $p_i$ 
on $\mathbf{F}=\mathbf{H}(\mathbf{\Phi},\mathbf{\Phi}^*, \mathbf{p})$ is localized. Consequently, only the field values $\mathbf{\Phi}$ and $\mathbf{\Lambda}$ 
corresponding to non-zero entries of $\Big(\frac{\partial \mathbf{F}}{\partial \mathbf{p}}\Big)$ are needed. Crucially, Eqs.~(\ref{eq:nladjoint}) 
and~(\ref{eq:nlgrad}) indicate that, no matter how many parameters $\mathbf{p}$, the full gradient $d g/d\mathbf{p}$ can be obtained with 
information from two field computations (one forward and one adjoint). 

While the above scheme can be performed in-silico with a significant computational cost reduction, it remains susceptible to simulation-reality 
gap, which can compromise the reliability of model-computed sensitivities when deployed in the physical system. A physics-informed scheme that minimizes 
reliance on accurate physics-based models for sensitivity evaluation is therefore desirable. The central obstacle to measuring $\mathbf{\Lambda}$ directly 
is that the adjoint field does not obey the same equation as the forward wave. This contrasts with reciprocal scenarios \cite{PSHPBWMMA2023,GWLK2025}. 
Instead, here, it propagates according to a linearized equation 
whose local response is proportional to the derivative of the forward nonlinearity, a condition that is not naturally realized for a generic nonlinear 
medium.

We resolve the challenge of emulating Eq. (\ref{eq:nladjoint}) within the same physical setting by using a pump-probe configuration under carefully 
designed adjoint sources. 
This approach exploits the linearization of the nonlinear system about the current operating point: the ``strong'' forward signal acts as the pump 
creating the local operating point of the nonlinear response, while a ``weak'' signal, sent through the same medium, acts as the probe. Indeed, suppose 
we slightly perturb the driving source and the field: $\mathbf{b} \to \mathbf{b} + \delta \mathbf{b}$ and $\mathbf{\Phi} \to \mathbf{\Phi} + \delta 
\mathbf{\Phi}$. To first order, the nonlinear equation $\mathbf{F}(\mathbf{\Phi,\Phi^*,p})=0$ reads:
\begin{align}
\Big(\mathbf{H}_{\text L}+[f({\cal I}_{n_0})+{\cal I}_{n_0}f'({\cal I}_{n_0})]\mathbf{P}_{n_0}\Big)\,\delta \mathbf{\Phi} \;+\; 
f'({\cal I}_{n_0})\Phi_{n_0}^2\mathbf{P}_{n_0}\,
\delta \mathbf{\Phi}^* \;\approx\; \delta \mathbf{b}~,
\end{align}
which emulates the adjoint Eq.~(\ref{eq:nladjoint}) with $\delta \mathbf{\Phi}^* \approx \epsilon\,\mathbf{\Lambda}^*$  in case of (i) perturbation 
sources such that $\delta \mathbf{b} = -\,\epsilon \,(\partial g/\partial \mathbf{\Phi})$, for $\epsilon \to 0$; and (ii) $f'({\cal I}_{n_0})\Phi_{n_0}^2$ 
being real-valued. Condition (i) is straightforward to implement experimentally. The central new element of our protocol is condition (ii), which 
we enforce by engineering the global phase of the source so as to impose the required field invariance under phase conjugation at the nonlinear defect.
Specifically, exploiting the global phase invariance of the wave equations (i.e., $e^{-i\theta^\star}\mathbf{\Phi}$ is the response for a phase-shifted 
injected wave $e^{-i\theta^\star}\mathbf{b}$), we choose $\theta^\star$ so that $e^{-2i\theta^\star}\,f'({\cal I}_{n_0}) \Phi_{n_0}^2 $ is real. In practice, 
we measure the phase of the steady-state field $\Phi_{n_0}$ at the nonlinear element and set $\theta^\star = \frac{1}{2}\arg\!\Big[f'({\cal I}_{n_0})
\Phi_{n_0}^2\Big]$. By rotating all fields by $e^{-i\theta^\star}$, we ensure that the linearized response has the same structure as the adjoint 
Eq. (\ref{eq:nladjoint}), which is crucial for extracting $\mathbf{\Lambda}$ from measurement.

We now summarize the steps of the proposed iNAP protocol for evaluating gradients, see Fig. \ref{Fig1}b. (i) Forward measurement: We excite the system 
with the driving source $\mathbf{b}$ and measure the steady-state field $\mathbf{\Phi}$ at positions associated to non-zero entries of $\partial 
\mathbf{F}/\partial \mathbf{p}$ and at the position of the nonlinear element. From this measurement, we extract the amplitude and phase 
of $\Phi_{n_0}$ to determine the phase-offset $\theta^\star$. (ii) Adjoint measurement: We re-inject the source with a phase shift $e^{-i\theta^\star}$ 
and superpose a small perturbation signal proportional to the field-gradient of the objective function. The adjoint source becomes
\begin{align}
\mathbf{b}_{\text{adj}} \;=\; e^{-i\theta^\star}\,\mathbf{b} \;-\; \epsilon\,e^{i\theta^\star}\,\big(\partial g/\partial \mathbf{\Phi}\big)^\top.
\label{adjsource}
\end{align}
We measure the resulting field, denoted $\mathbf{\Psi}$, under the driving excitation 
$\mathbf{b}_{\text{adj}}$. (iii) Adjoint field evaluation: We compute the adjoint field as
\begin{align}
\mathbf{\Lambda} \;=\; \frac{\mathbf{\Psi} - e^{-i\theta^\star}\mathbf{\Phi}}{\epsilon};\quad \epsilon\rightarrow 0.
\label{lambda}
\end{align}
In practice, $\epsilon$ is chosen so that the probe remains in the linear-response regime around the pumped operating point while remaining large 
enough for accurate phase measurement. (iv) Gradient assembly: We compute $dg/d\mathbf{p}$ using Eq.~(\ref{eq:nlgrad}). 

Notably, our iNAP provides the full gradient information without needing a numerical solution of the nonlinear equation $\mathbf{F}(\mathbf{\Phi},
\mathbf{\Phi}^*,\mathbf{p})=0$ or evaluating the entire Jacobian term-by-term. Instead, the physical system itself performs these operations while 
inherently accounting  for all system complexities, including losses and detunings. The iNAP approach also circumvents the prohibitively large 
memory requirements of brute-force adjoint simulations by directly accessing the steady-state response of the nonlinear system.  Once $d g/d\mathbf{p}$ 
is known, we can employ any gradient-based optimization algorithm to adjust the parameters and improve the objective. We set up an external control 
enclosure to orchestrate the entire process (see Fig. \ref{Fig1}c), including the sequential (forward and adjoint) wave-field excitations, in-situ 
measurements, gradient computations, and optimization updates, ensuring seamless and efficient real-time optimization. This comprehensive and physically 
faithful representation of the system’s response enables highly precise and reliable optimization outcomes, opening the door to real-time control 
of nonlinear wave phenomena.

\subsection*{Physical Platform and Implementation of iNAP}

We chose to demonstrate the validity of our iNAP using a prototype platform of complex multipath networks of coaxial cables that describes systems with 
wave chaotic dynamics \cite{kottos2000,kottos2001}. Such a platform has been successfully used for the 
description of mesoscopic quantum transport, as well as for sound and electromagnetic wave propagation in complex interconnected structures such as buildings, 
ships, and aircrafts \cite{kuchment2004,Baum1986}. The networks consist of $n=1,\dots,V$ vertices (T-junctions) connected by 
one-dimensional coaxial cables (bonds) $B=(n,m)$ of length $L_{B}$, chosen to be mutually incommensurate. The position $x_B$ on bond $B$ is measured from 
the $n$-th vertex ($x_B=0$) to the $m$-th vertex ($x_B=L_B$). The connectivity of the network is described by the $V\times V$ symmetric adjacency matrix 
$\mathcal{A}$ with elements $\mathcal{A}_{nm}=1$ when vertices $n\neq m$ are connected by a bond and $\mathcal{A}_{nm}=0$ otherwise. The voltage between 
the inner and outer conductor surfaces of the coax cables at position $x$ along each bond, satisfies the telegrapher equation with wavevector $k=\omega 
n_r/c$, where $c$ is the speed of light in vacuum and $n_r$ is the complex refractive index of the coaxial cables; its imaginary part accounts for Ohmic 
losses (see Methods). Solving the bond equations together with the continuity and current-conservation conditions at the network vertices leads to a reduced 
nodal description in terms of the vertex fields $\Phi_n$, defined by the voltages at the $n$-th vertex.

Without loss of generality, we assume that the nonlinearity is located at the $V$-th vertex. Experimentally, this nonlinear vertex is implemented by replacing 
the corresponding T-junction with a dielectric cylindrical resonator inductively coupled to a metallic ring short-circuited by a Schottky diode 
positioned above the resonator (see Methods). The resonator is coupled to the surrounding network through three kink antennas placed at the 
endpoints of three coaxial cables with resonator-antenna coupling strengths $\gamma_1,\gamma_2,\gamma_3$. The resonator, the ring diode, and the 
kink antennas are arranged inside a cylindrical cavity that confines the electromagnetic field and thereby enables a strong localized 
nonlinear response around $\nu=6.382\,\mathrm{GHz}$, see Fig. \ref{Fig2}a. We describe this element within a temporal coupled mode theory 
(TCMT), with a single complex mode amplitude $a$, normalized such that $|a|^2$ is the energy stored in the resonator mode. The diode-induced 
response is modeled by a saturable nonlinearity $f(|a|^2)=\frac{4\pi z_1}{1+\chi |a|^2}$ while the linear impedance of the resonator is $(H_L)_{VV}
=-4\pi z_0-\sum_{i=1}^{3}\gamma_i^2\cot(kL_i)$, describing both its intrinsic linear response and its loading through the kink-antenna couplings. 
The complex parameters $z_0,z_1,\chi$ together with the coupling strengths $\gamma_1,\gamma_2,\gamma_3$ were extracted from direct fits 
to the measured scattering parameters of the nonlinear vertex (see Methods). 

We describe the whole platform within a hybrid network-TCMT framework (see Methods and Supplementary Information): the linear background 
network is treated by a graph formalism \cite{kottos2000}, while the nonlinear resonator by TCMT. Within this formalism, the resonator 
is treated explicitly as the $V$-th node of the reduced system, so that the state vector that characterizes a steady state of the setup is $\mathbf{\Phi}
=(\Phi_1,\Phi_2,\cdots,\Phi_{V-1},a)^\top$. 

We convert the compact network into an open scattering system by attaching transmission lines (TLs) to $M<V$ vertices, which inject and collect 
monochromatic waves of angular frequency $\omega=2\pi \nu$. Their coupling to the graph is described by the $M\times V$ matrix $\mathbf{W}$, 
with entries $W_{\alpha n}=1$ if TL $\alpha$ is attached to vertex $n$, and $W_{\alpha n}=0$ otherwise. An incident wavefront $\mathbf{I}=
(I_1,\ldots,I_M)^\top$, with $I_\alpha=A_\alpha e^{i\theta_\alpha}$, drives the system through the source term $\mathbf{b}=2i\mathbf{W}^\top
\mathbf{I}$. The steady-state equation follows by imposing continuity and current-conservation conditions at the vertices. In the hybrid formulation, 
the steady-state scattering problem (neglecting higher harmonics that are not excited in our setting) is written compactly as
\begin{equation}
\mathbf{F}\!\left(\mathbf{\Phi},\mathbf{\Phi}^\ast, \mathbf{p}\right)=
\left(
\begin{bmatrix}
\mathbf{H}_{\rm L}^{(V-1)} & \mathbf{G}\\
\mathbf{G}^\top & (H_{\rm L})_{VV}
\end{bmatrix}
+i\mathbf{W}^\top\mathbf{W}
\right)\mathbf{\Phi}
+f(|a|^2)\,\mathbf{P}_V\mathbf{\Phi}
-\mathbf{b}
=0,
\label{GSE}
\end{equation}
where the term inside the parentheses represents $\mathbf{H}_{\text L}$. Here $(P_V)_{nm}=\delta_{nm}\delta_{mV}$, while 
$\mathbf{H}_{\rm L}^{(V-1)}$ is the $(V-1)\times(V-1)$-dimensional graph operator of the linear subnetwork with matrix elements 
\begin{equation}
(H_L^{(V-1)})_{nm}=
\begin{cases}
-\displaystyle\sum_{\ell\neq n}^{V-1} \mathcal{A}_{n\ell}\cot(kL_{n\ell}), & n=m,\\[1.2ex]
\mathcal{A}_{nm}\csc(kL_{nm}), & n\neq m,
\end{cases}
\qquad n,m=1,\ldots,V-1.
\label{MGE}
\end{equation}
Finally, the $V-1$-dimensional vector $\mathbf{G}$ encodes the coupling of the resonator to the three kink-antenna endpoints; its 
components are $G_{l}=\gamma_{l_r}\,\csc(kL_{l_r}) \delta_{l,l_r}$, where $L_{l_r}$ is the length of the $l_r$-th coupled coaxial segment. 

The gradient sensitivities are evaluated using Eq. (\ref{eq:nlgrad}). The implementation of this equation requires the knowledge of 
the adjoint field $\mathbf{\Lambda}$. For this process, the design of the appropriate driving source requires the evaluation of 
$\partial g/\partial \mathbf{\Phi}$, which is obtained from the specific form of $g(\mathbf{\Phi},\mathbf{\Phi}^*,\mathbf{b})$.

The evaluation of Eq. (\ref{eq:nlgrad}) requires also knowledge of the gradient $\partial \mathbf{F}/\partial\mathbf{p}=(\partial\mathbf{H}/
\partial\mathbf{p})\mathbf{\Phi}-\partial\mathbf{b}/\partial\mathbf{p}$. The optimization vector $\mathbf{p}$ incorporates cavity-shaping  
parameters (e.g. selected set of network bond lengths $\{L_{nm}^{\rm opt}\}$), and/or the amplitudes $A_\alpha$ and phases $\theta_\alpha$ of 
the injected waves from the $\alpha-$th TL. The former parameter set is encoded in $\mathbf{H}_{\rm L}$. Its gradient $\frac{\partial 
{\mathbf H}_{\rm L}}{\partial L_{nm}^{\rm opt}}$ is a sparse $V \times V$ operator with non-zero elements {\it only} at entries that 
depend on the selected bonds $\{L_{nm}^{\rm opt}\}$. The optimization parameters associated with the injected waves are encoded in {\bf b}; resulting in 
${\displaystyle \frac{\partial b_n}{\partial A_\alpha} = 2 i e^{i \theta_\alpha} W_{\alpha,n}}$, and ${\displaystyle \frac{\partial b_n}{\partial 
\theta_\alpha} = -2 A_\alpha e^{i \theta_\alpha} W_{\alpha,n}}$.

Eventually, the objective function gradient becomes:
\begin{equation}
    \frac{d g}{d \mathbf{p}} \equiv \left[ \frac{d g}{d L_{nm}^{\rm opt}},\ \frac{d g}{d A_\alpha},\ \frac{d g}{d \theta_\alpha} \right]=
\begin{bmatrix}
        \displaystyle 2 \mathcal{R} \left\{ \mathbf{\Lambda}^\top \frac{\partial {\mathbf H}_L}{\partial L_{nm}^{\rm opt}} \mathbf{\Phi} \right\} \\[2ex]
        \displaystyle \frac{\partial g}{\partial A_\alpha} + 2 \mathcal{R} \left\{ \mathbf{\Lambda}^\top \frac{\partial \mathbf{b}}{\partial A_\alpha} \right\} \\[2ex]
        \displaystyle \frac{\partial g}{\partial \theta_\alpha} + 2 \mathcal{R} \left\{ \mathbf{\Lambda}^\top \frac{\partial \mathbf{b}}{\partial \theta_\alpha} \right\}
    \end{bmatrix}^\top.
    \label{grad_graph}
\end{equation}
Equation (\ref{grad_graph}) highlights the reduced measurement complexity required for gradient evaluation. Specifically, because the controllable parameters 
affect only selected entries of ${\mathbf H}_{\rm L}$ and ${\mathbf b}$, only the corresponding components of the fields $\mathbf{\Phi}$ and $\mathbf{\Lambda}$ associated 
with the non-zero entries of ${\partial{\mathbf H}_{\rm L}\over \partial L_{nm}^{\rm opt}}$ and ${\partial {\bf b}\over \partial A_\alpha}$, ${\partial {\bf b}\over 
\partial \theta_\alpha}$ need to be measured. 

\subsection*{Examples of Modalities and Their Optimization Objectives}

Below, we define representative objective functions for in-situ optimization of various nonlinear wave-control tasks. Importantly, because the 
response is nonlinear, the input intensity itself becomes an additional control knob, enabling reconfigurable multiport functionalities beyond 
the linear regime.

{\it Arbitrary Power Splitting $(P_1^{\rm tar}\!:\!P_2^{\rm tar}\!:\!\cdots\!:\!P_{M_{\rm tar}})$ --} To illustrate on-the-fly adaptability 
of the iNAP to various objectives, we consider the task of enforcing a prescribed distribution of power $\{P_\alpha^{\rm tar}\}_{m=1}^{M_{\rm tar}}$ 
among $M_{\rm tar}\leq M$ targeted ports. In a lossy nonlinear platform, the relevant target is not the 
absolute throughput, but the fraction of the available transmitted power directed to each channel. Denoting by $P_\alpha$ the output power 
at targeted port $\alpha$, we quantify the mismatch from a desired output profile $\{P_\alpha^{\rm tar}\}_{m=1}^{M_{\rm tar}}$, through
\begin{equation}
  g_{\rm PS} \;=\;
  \sum_{\alpha'=1}^{M_{\rm tar}}\left|\frac{P_{\alpha'}}{\sum_{\alpha=1}^M P_\alpha} - P_{\alpha'}^{\rm tar}\right|;\quad {\rm with} 
  \quad\sum_{\alpha=1}^{M_{\rm tar}}P_m^{\rm tar}=1.
  \label{eq:gapr}
\end{equation}
This objective penalizes deviations from the prescribed redistribution of power independently of the overall transmission level and is 
therefore well suited to dissipative settings. It is also agnostic to the mechanism by which the target partition is 
realized, i.e., whether through shaping of the incident wavefront, modification of the effective cavity response, or both. Perfect power 
splitting corresponds to $g_{\rm PS}=0$.

{\it Nonlinear Coherent Perfect Absorption --}In linear media, coherent perfect absorption (CPA) arises when a properly phased multi-port 
excitation drives a (weakly) absorbing cavity into a state with no outgoing radiation, owing to interference that traps the incident energy 
until it is fully absorbed \cite{WCGNSC2011,CKA2020,FSK2017,aluCPA}. Nonlinearity introduces an additional 
handle for controlling CPA through the input intensity itself, thereby enriching the accessible absorption states and their spectral 
characteristics. At the same time, this added freedom is accompanied by the loss of superposition, which makes prediction and design considerably 
more challenging. New computational approaches have recently been developed to overcome these difficulties, and experiments have confirmed 
their feasibility \cite{WGTKK24,Galiffi2026}. The iNAP can be utilized for the management of the multi-path interference via 
cavity and/or wavefront shaping. The optimization objective function is
\begin{equation}
    g_{\text{CPA}} = 1 - \frac{\sum_{\{I_\alpha\}} |\phi_\alpha - A_\alpha e^{i \theta_\alpha}|^2}{\sum_{\{I_\alpha\}} |A_\alpha|^2} - 
    \frac{\sum_{\{T_\beta\}} |\phi_\beta|^2}{\sum_{\{I_\alpha\}} |A_\alpha|^2}
\end{equation}
measuring residual outgoing energy in all channels. Specifically, the second term describes the reflected waves 
from the injected channels $\{I_\alpha\}$ and the third term describes the transmitted waves from the remaining $\{T_\beta\}\neq \{I_\alpha\}$ 
channels; normalized by the total input power. Perfect absorption corresponds to $g_{\text{CPA}}=1$.

{\it Invisibility --} Suppressing the scattering signature of a target entails erasing the phase and amplitude signatures left on interrogating 
waves after they engaged with the target. In our setting, this is achieved by programming the multipath interference inside the scattering 
region, either through cavity shaping or through suitable control signals that compensate both elastic and absorptive scattering imprints. To 
quantify this goal, we define \cite{GWLK2025}
\begin{equation}
    g_{\text{invis}} = \frac{|\phi_{\alpha_0} - A_{\beta_0} e^{i \theta_{\beta_0}}|^2}{A_{\beta_0}^2}+  
    \frac{|\phi_{\beta_0} - A_{\beta_0} e^{i\theta_{\beta_0}}|^2}{\sum_{\{I_\beta\}} |A_\beta|^2}+ 
    \frac{\sum_{\{T_{\alpha\neq \alpha_0,\alpha_c}\}} |\phi_\alpha|^2}{\sum_{\{I_\beta\}} |A_\beta|^2},
    \label{transparent}
\end{equation}
where $g_{\rm invis}=0$ indicates optimal invisibility performance. Above, the first term on the right-hand-side enforces matching between the 
outgoing field from the probed channel $\alpha_0$ and the reference signal injected through channel $\beta_0$, thereby suppressing detectable 
phase and amplitude distortions. The second term penalizes reflection back into the interrogation channel $\beta_0$, while the third suppresses 
leakage into all remaining output channels except the probe channel $\alpha_0$ and the control channel $\alpha_c$ (which was left unconstrained).

{\it Asymmetric Transport --} A hallmark of nonlinear wave transport is the emergence of direction-dependent transmission under otherwise mirrored 
excitation conditions. To quantify this effect, we consider two experiments performed at the same frequency and input power: in the first, port 1 
is excited, and the transmitted power is measured at port 2; in the second, port 2 is excited under identical conditions and the transmitted power 
is measured at port 1. We then define
\begin{equation}
  g_{\mathrm{Asym}} \;=\;
  \frac{|\phi_{2}^{(1)}|^2}{|\phi_{1}^{(2)}|^2}\,,
  \label{eq:gasym}
\end{equation}
where the superscript indicates the driven port. In linear, time-invariant systems, reciprocity dictates that $g_{\rm Asym}=1$. In nonlinear systems 
with structural asymmetry, it can deviate substantially from one, thereby quantifying diode-like transmission. 

\subsection*{In-situ Implementation of iNAP}
\begin{figure}[htbp]
  \centering
  \includegraphics[width=0.775\linewidth]{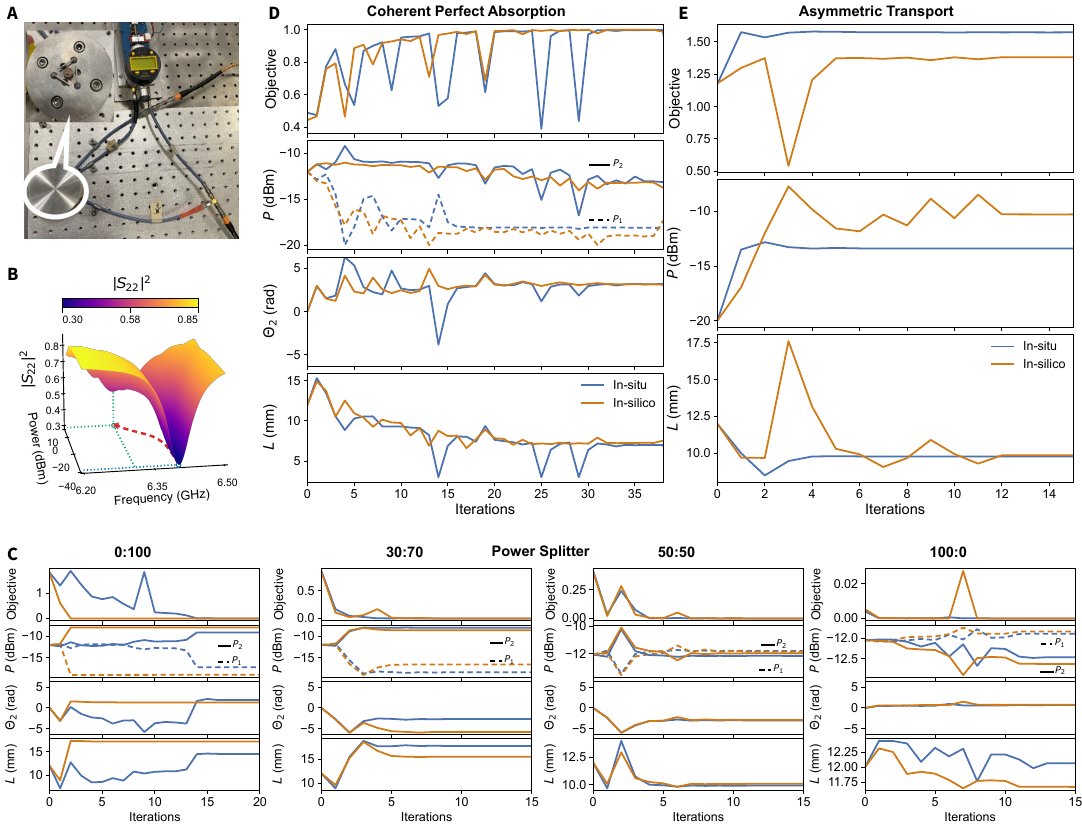}
  \vspace{-6mm}
  \caption{{\bf Experimental implementation of the iNAP protocol}. {\bf (A)} The physical platform consisting of coax cables arranged in a loop 
  geometry with one of the two vertices being nonlinear. The electrical length of one arm of the loop is digitally controlled 
  via a phase-shifter. The nonlinear vertex (left) consists 
of a cylindrical cavity that hosts a dielectric resonator that is inductively coupled to a ring antenna that is short-circuited by a Schottky 
diode (inset); {\bf (B)} The reflectance $|S_{11}|^2$ of the nonlinear cavity versus frequency and power of the injected wave. The nonlinear 
response is enhanced around $\nu\approx 6.382$GHz; {\bf (C)} Optimization progress for {\it Arbitrary Power Splitting} targeting various relative 
output power profiles: $(P_1^{\rm tar}\!:\!P_2^{\rm tar})=(0{:}100)$ (first column), $(P_1^{\rm tar}\!:\!P_2^{\rm tar})=(30{:}70)$ (second column), 
$(P_1^{\rm tar}\!:\!P_2^{\rm tar})=(50{:}50)$ (third column), and $(P_1^{\rm tar}\!:\!P_2^{\rm tar})=(100{:}0)$ (fourth column), where 
$P_{1,2}=A_{1,2}^2$. Upper rows report the convergence process of the objective function Eq. (\ref{eq:gapr}) versus the number of iterations. 
Second, third, and fourth rows report the evolution of the optimization parameters $\big(P_1,\,P_2,\,\theta,\,L_{12}^{\rm opt}\big)$, towards 
their convergence value. {\bf (D)} Optimization progress for nonlinear {\it Coherent Perfect Absorption}. 
Top panel: The convergence of $g_{\text{CPA}}$ vs.\ iteration, towards a maximum absorption of $0.998$, with an over 70 dB difference between input 
and output power; The other three panels report the convergence of the optimization parameters $\big(P_1,\,P_2,\,\theta,\,
L_{12}^{\rm opt}\big)$, respectively, towards their optimal value. {\bf (E)} Optimization progress of the objective function $g_{\mathrm{Asym}}$ for {\it Asymmetric 
Transport}. The optimizer converged in only a few iterations, resulting in a power difference of roughly 60\% favoring port 1 (upper row). Second and third 
row report the evolution of the optimization parameters $\big(P,\,L_{12}^{\rm opt}\big)$, towards their convergence value. In 
{\bf C,D,E}, the in-situ (in-silico) results are indicated with blue (orange) lines.} 
  \label{Fig2}
\end{figure}

For the in-situ demonstration of iNAP, we used a minimal nonlinear multipath microwave network consisting of two incommensurate coax cables connected 
through two vertices: a nonlinear vertex on one side and a standard T-junction on the other, thereby forming a loop (see Fig. \ref{Fig2}a). A TL is 
attached to each vertex. A digitally controlled phase shifter is inserted on one of the bonds, providing a tunable electrical length $L_{12}^{\rm opt}$ 
for ``cavity-shaping'' purposes. The system is instrumented with a two-source VNA. The optimization process occurred at a fixed operational frequency 
$\nu\approx 6.38$GHz. In this range, the response of the nonlinear vertex acquires its maximum value, see 
Fig. \ref{Fig2}b. Using the same hardware, we have further tested the validity of the iNAP for three modalities, namely, arbitrary power splitting, 
nonlinear CPA, and asymmetric transport.

For the coherent two-port modalities (arbitrary power splitting and CPA), the optimization parameters have been chosen to be ${\mathbf p}=
\big(A_1,\,A_2,\,\theta,\,L_{12}^{\rm opt}\big)$, where $A_{1,2}$ are the drive amplitudes on TLs $\alpha=1$ and $\alpha=2$, and $\theta$ is their 
relative phase (with respect to TL $\alpha=1$). For asymmetric transport (AT), which uses a single coherent 
input, we optimized only two parameters, ${\mathbf p}=\big(A,\,L_{12}^{\rm opt}\big)$ where $A$ is the active-port drive level.

In the upper row of Fig. \ref{Fig2}c, we report the in-situ results (blue lines) of the optimization objective function for a variety of output power 
splitting profiles between the two outgoing channels. Using the {\it same hardware}, we have achieved on-the-fly arbitrary relative output power profiles 
$P_1^{\rm tar}=0$ and $P_2^{\rm tar}=1$ (first column), $P_1^{\rm tar}=0.3$ and $P_2^{\rm tar}=0.7$ (second column), $P_1^{\rm tar}=P_2^{\rm tar}=0.5$ 
(third column), and $P_1^{\rm tar}=1$ with $P_2^{\rm tar}=0$ (last column). The convergence of the four optimization parameters towards their optimal 
value for each of the four cases is reported in the second, 
third, and fourth rows of these subfigures, respectively. Similarly, the upper panel of Fig.~\ref{Fig2}d shows the experimental results for $g_{\rm 
CPA}$ vs. iteration number in case of a nonlinear CPA. The evolution of the four optimization parameters with the iteration number is shown in the 
rows below. Finally, in the upper panel of Fig. \ref{Fig2}e we report the experimental results for the optimization objective function $g_{\rm Asym}$ 
for asymmetric transport. At the same figures, we also report the results of the in-silico implementation (orange lines) of the iNAP obtained from a 
digital twin that models the network. The differences between the converged in-situ and in-silico optimization parameters likely arise from convergence 
to distinct local minima in the multivalued parameter landscape. Despite this, the two approaches attain nearly identical asymptotic objective values.

\subsection*{Scale-up in-silico Validation of iNAP using Large Multipath Networks}

\begin{figure}[htbp]
  \centering
  \includegraphics[width=0.75\linewidth]{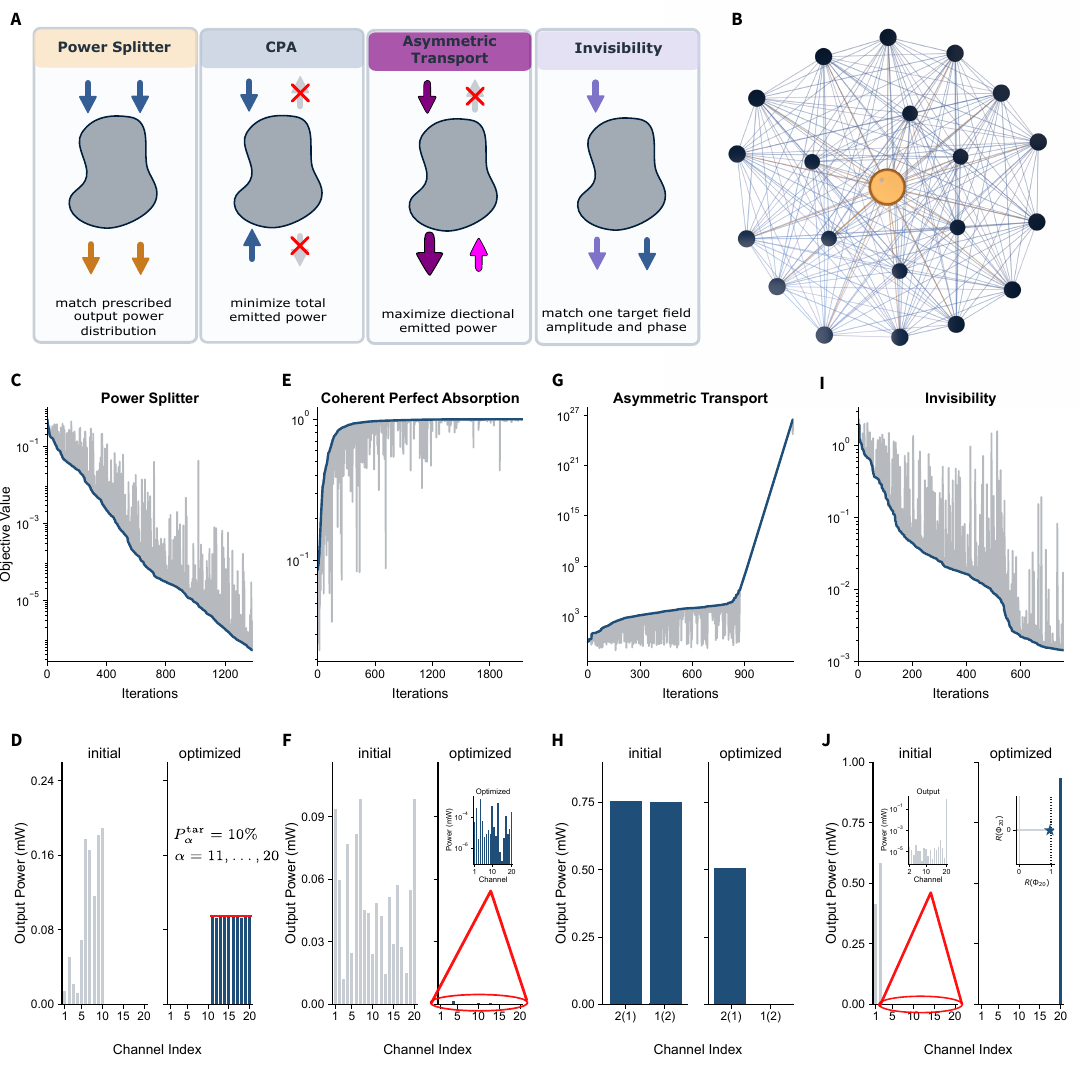}
\vspace{-10mm}
  \caption{{\bf In-silico validation of iNAP in a large-scale network.} {\bf (A)} Testing iNAP across multiple modalities in a complex scattering 
  setup. {\bf (B)} The in-silico network consists of $V=21$ vertices, one of which hosts a nonlinear defect. Owing to its multiresonant topology, 
  the network supports a combinatorially large number of interfering multipath scattering trajectories, enabling us to test iNAP in a highly complex 
  wave environment. {\bf (C)} Progression of objective function $g_{\rm PS}$ for Arbitrary Power Splitter targeting an equidistributed outgoing power 
  profile among $\alpha=11,\cdots, M=20$ channels and zero at all others. {\bf (D)} Relative output power profile at the end of the first (left subfigure) 
  and last (right subfigure) iteration. {\bf (E)} Progression of objective function $g_{\rm CPA}$ for nonlinear CPA and 
  {\bf (F)} the output power profile at the end of the first (left subfigure) and last (right subfigure) iteration for a network corresponding to total 
  absorption $2.4\% (\approx 100\%)$. The inset in the right subfigure is a magnification of the outgoing power at the channels. {\bf (G)} Progression 
  of objective function $g_{\rm Asym}$ for asymmetric transport in case of $M=2$ TLs {\bf (H)} and the output power profile at the end of the first (left subfigure) 
  and last (right subfigure) iteration. {\bf (I)} Progression of objective function $g_{\rm invis}$ for invisibility in case of $M=20$ channels {\bf (J)} 
  and the output profile at the end of the first (left subfigure) and last (right subfigure) iteration. In the inset of the left subfigure, we 
  report the outgoing powers from TLs $\alpha=2,\cdots, 20$. In the inset of the right subfigure, we report the phase difference between the input 
  and output signal at the end of the optimization process.} 
  \label{Fig3}
\end{figure}

The in-situ realization above provides a proof of principle for iNAP in hardware. We now extend the same framework to an in-silico implementation in a 
more complex multipath network, where the larger number of scattering pathways and control degrees of freedom provide a stringent test of its scalability, 
versatility and robustness. This setting allows us to probe how the protocol performs in regimes of enhanced interference complexity while preserving 
the same underlying nonlinear-scattering principles. We benchmark iNAP across a range of target functionalities, spanning the modalities illustrated in 
Fig. \ref{Fig3}a.

We considered fully connected networks consisting of $V=21$ vertices with a total of $210$ bonds, see Fig. \ref{Fig3}b. One of these vertices hosts a 
nonlinear resonator, with the same nonlinearity described in Section 3. The bond-lengths are initially uniformly distributed in the interval 
$[L_0-\delta L, L_0+\delta L]$ where $L_0=22$cm and $\delta L=20$cm. We have attached $M=20$ TL (one to each vertex, apart from the nonlinear 
one) that have been used for injecting (receiving) the interrogating (scattering) signal. The incident wavefronts were fixed and chosen to be random and 
were injected from the first 10 (all 20) TLs for the power splitter (CPA) application. The frequency of the injected waves was chosen to be $\nu=6.382$GHz 
for all cases and the Ohmic losses of the cables were modeled using the experimentally extracted value of the reflective index (see Methods). The 
iNAP optimization has been achieved via bond-length variations that aimed to maximize/minimize the predefined objective function.

Subfigures Fig. \ref{Fig3}c,d report the optimization results for an arbitrary power splitter. As an example case, we consider bond configurations 
that lead to a $10\%$ uniform relative outgoing power split among a selected subset of target channels, $\alpha=11,…,20$, and zero elsewhere. As 
shown in Fig. \ref{Fig3}c, the objective function decreases rapidly with iteration number and reaches values as low as $10^{-7}$, indicating agreement 
between the realized and target output distributions. Figure \ref{Fig3}d compares the relative output-power profile across all $M=20$ channels at 
the initial (left) and final (right) iterations of the iNAP protocol, highlighting the emergence of the desired uniform splitting over the selected 
output TLs.

Next, we demonstrate nonlinear CPA using iNAP. In Fig. \ref{Fig3}e, 
we show the convergence of the objective function towards a value $g_{\rm CPA}= 0.999$ occurring after 1800 iterations, while Fig. \ref{Fig3}f 
reports the output power after the first (left) and last (right) iteration of the iNAP protocol. In the former case, the total absorption is $A
\approx 0.024$ while the optimized bond configuration after the last iteration leads to (near) perfect absorption of $A\approx 0.999$.

We have also implemented iNAP for asymmetric transport between two channels $\alpha=1,2$.  In Fig. \ref{Fig3}g we show the convergence 
of the objective function $g_{\rm Asym}$ to an approximate value $10^{27}$. Subfigures Fig. \ref{Fig3}h shows the actual output power at TL 
$\alpha=2 (1)$ for incident waves from TL $\alpha=1 (2)$, at the first (left subfigure) and at the last (right subfigure) iteration of the optimization 
process. 

The last example reported in Fig. \ref{Fig3} shows the in-silico performance of iNAP for cavity camouflage (invisibility). Specifically, the 
bond configuration is optimized such that a wave that is injected through channel $\alpha=1$ with amplitude $A_{\beta_0}=1$ and phase 
$\theta_{1}=0$ will emerge from channel $\alpha=20$ with the same amplitude and phase as the incident wavefront. To compensate for network 
losses, an additional control signal is launched from channel $\alpha=2$ with amplitude $A_{2}=0.765$ and phase $\theta_{2}=-55.3^\circ$. Figure 
\ref{Fig3}i shows the progression of the objective function $g_{\rm invis}$ towards values as small as $g_{\rm invis}\approx 9\times 10^{-3}$ 
after $600$ iterations. Here, in addition to the constraints imposed by the objective function $g_{\rm invis}$ in Eq. (\ref{transparent}), 
we have also requested zero transmission and reflection from the injected and control channel $\alpha=2$, which amounts to the following 
modification of the optimization function $g_{\rm invis}\rightarrow g_{\rm invis}+\frac{|\phi_c-A_{\alpha_c}e^i\theta_{\alpha_c}|^2}
{\sum_{\{I_\beta\}} |A_\beta|^2}$. The left (right) subfigure of Fig. \ref{Fig3}j shows the power channel profile of the outgoing signal after 
the first (last) iteration of the iNAP protocol. In the former case, the outgoing power profile demonstrates a strong reflection at the injected channels due 
to direct processes associated with fully connected networks. In contrast, the outgoing wave at the last iteration of the iNAP protocol is extracted 
from the targeted channel $\alpha=20$ and has the same amplitude and phase as the incident wavefront.

\section*{Discussion}
We have introduced a route to in-situ optimization in non-reciprocal complex wave platforms operating under nonlinear, multipath, scattering conditions. 
In such settings the challenge is not only the breakdown of the superposition principle but most importantly the fact that in-situ back-propagation becomes 
unrealizable within the same hardware using any existing protocols. Rather than treating multiple scattering, complex interference, and the breakdown of 
superposition as obstacles, iNAP harnesses them directly by judiciously engineered adjoint excitations so that the same physical 
system can carry out both the forward and adjoint operations. This allows the evaluation of the gradient by measuring the excitation fields at selective 
positions, independent of the dimensionality of the control space. In this way, the protocol largely sidesteps the simulation–reality gap that limits 
purely in-silico optimization and replaces digital gradient estimation with a measurement-driven procedure executed in hardware. At the same time 
iNAP applies to any type of nonlinearity, while it self-calibrates to losses and detunings.

Our proof-of-principle experimental demonstration establishes the physical viability of iNAP for real-time control of electromagnetic-wave functionalities 
in nonlinear, multiresonant and multipath settings, despite loss, backscattering and other imperfections that are usually regarded as detrimental to 
existing physical computing schemes \cite{GBWL21,SGL25,hughes2018,PSHPBWMMA2023,LM23,DMW2025,GWLK2025}. The complementary in-silico implementation in 
a highly complex network shows that the same strategy is scalable and becomes highly effective when more complex degrees of freedom are engaged, resulting 
in greater training fidelity across a range of target functionalities. Taken together, these results indicate that iNAP is not tied to a particular geometry 
or objective, but instead provides a general approach to programming nonlinear scattering systems through direct physical access to their gradients.
More broadly, the platform can adapt on the fly to changing objectives, with the full optimization loop - sensing, backpropagation, and actuation - executed 
within the physical hardware itself rather than through digital simulation. 

Looking ahead, iNAP suggests a route to adaptive wave systems operating in partially unknown environments without requiring a digital twin. In communications, 
this could enable self-optimizing routing and power distribution in complex media; in sensing, it could support operating points tailored in-situ for enhanced 
selectivity, asymmetry or absorption; and in analog physical intelligence, it offers a route by which strongly scattering nonlinear hardware can be trained 
directly through its own dynamics. Important next steps will be to extend the protocol to platforms with distributed nonlinearities, gain and non-conservative 
elements, or explicit microscopic time-reversal-symmetry breaking, where the adjoint structure becomes richer and the computational opportunities broader. 

Our approach differs fundamentally from conventional digital and neuromorphic AI frameworks, where training and adaptation are typically mediated by an 
external computational layer. This points to wave platforms that do not merely process signals, but learn and reconfigure directly through their own 
scattering dynamics.

\section*{Materials and Methods}

\subsection*{Experimental Platform}
The loop-network used in our experiments (see Fig. \ref{Fig2}a) consists of two bonds made by coaxial cables (Huber+Suhner S04272) of length $L_1$ 
and $L_2$ connected at one end at the nonlinear vertex via kink antennas and at the other end at the ports of a T-junction. The first bond has length 
$L_1\approx 28.7$ cm. The second bond consists of two coax-cables of length $L_2^{(1)}\approx 31.9$ cm and $L_2^{(2)}\approx 26.6$ cm connected to a 
digitally controlled phase shifter. Its total length is $L_2=L_2^{(1)}+L_2^{(2)}+\delta$ where $\delta\in[2,42]$ cm being the length variation performed 
by the phase shifter. A VNA is connected to the remaining 
port of the T-junction and to a coax cable of length $L_3=22.7$ cm that has been coupled to the nonlinear resonator via a kink antenna. The Ohmic losses 
of the cables are encoded in the imaginary part of their refractive index, which could be obtained via a best fit of the measured frequency-dependent 
transmission [ $t(\omega) = e^{i\omega L/c(n_r+in_i)}$] through a cable of a specific length. Best fitting analysis indicated that $n_i\approx 2 
\times 10^{-3}$, whereas $n_r\approx 1.212$. 

\subsection*{Experimental Implementation and Characterization of the Nonlinear Vertex}
In the experiment, the nonlinear vertex is implemented using a cylindrical ceramic resonator (ZrSnTiO, $\varepsilon\approx 36$, height 5mm, diameter 
8mm), with resonance frequency $\nu_0\approx 6.885$GHz and linewidth $\gamma\approx 1.7$MHz; see Figs.~\ref{Fig2}a. The resonator is inductively coupled 
to a 3mm-diameter metallic ring short-circuited by a Schottky detector diode (SMS 7630-079LF, Skyworks). The resonant $z$-directed magnetic field 
induces a current in the ring, generating a voltage across the diode whose magnitude depends on the incident power. Low input power leaves the device 
in a weakly nonlinear regime, whereas higher power drives it into a strongly nonlinear one. The resonator is coupled to the network and an attached 
transmission line through kink antennas. The whole structure is embedded in an aluminum cylindrical cavity (diameter $\sim 15$ cm, height $\sim 10$ cm), 
which further enhanced the nonlinear response. The nonlinear vertex is designed to operate in the 6.36–6.5GHz range and has already been used for 
the analysis of nonlinear CPAs \cite{WGTKK24} and for asymmetric transport \cite{WKKK23}.

We characterize the nonlinear vertex using temporal coupled mode theory (TCMT) \cite{WKKK23}. The resonator couples to the three kink antennas through 
$W_R=(\gamma_1,\gamma_2,\gamma_3)$, while direct processes between antennas are neglected. Its dynamics are described by
\begin{align}
i\frac{d}{dt}a(t) &= \left(\tilde{\omega}-i\frac{W_RW_R^\top}{2}\right)a(t) + iW_R|S_+\rangle, \nonumber\\
|S_-\rangle &= -|S_+\rangle + W_R^\top a(t).
\label{oneres}
\end{align}
where $a(t)$ is the amplitude of the magnetic field of the resonator mode and $|S_\pm\rangle$ are the incoming and outgoing channel amplitudes. The effective 
resonance angular frequency $\tilde{\omega} =  2\pi (z_0 - \frac{z_1}{1+\chi_s |a|^2})$ contains the diode-induced saturable contribution, which depends on 
the resonator intensity $|a|^2$. Under monochromatic excitation $|S_\pm\rangle=S_\pm e^{-i\omega t}$ and 
$a(t)=ae^{-i\omega t}$ (experimentally, higher harmonics are negligible, so the response is well described at the driving frequency alone).  
Substitution of the temporal form of the fields in Eq. (\ref{oneres}) leads to the following equations for the field amplitudes,
\begin{equation}
\omega a = \left(\tilde{\omega}-i\frac{W_RW_R^\top}{2}\right)a + iW_RS_+, \quad\quad S_- = -S_+ + W_R^\top a
\label{Eq:CMT-2}
\end{equation}
from which the nonlinear scattering matrix follows as
\begin{align}
S = -1 + iW_R^\top\frac{1}{\omega - \left(\omega_0 + 2\pi z_0 - \frac{2\pi z_1}{1+\chi_s |a|^2}\right)+i\frac{W_RW_R^\top}{2}}W_R
\label{Sres}
\end{align}
which is used for extracting the fitting parameters via comparison with our measurements, see Ref. \cite{WKKK23}. Our best fit between the theoretical 
expression Eq. (\ref{Sres}) and the measured $S$-parameters results in the following values $z_0 = (-86.4 - i\cdot 59.2)$ MHz, $z_1 = (-86.4 - i\cdot 50.0)$ MHz, $\gamma_1^2=\gamma_2^2=\gamma_3^2 = 62.5$ MHz, $\chi_s = (1.5+i)\cdot 10^{9}$(mW$\cdot$s)$^{-1}$. 

\subsection*{Phase shifter calibration}

To tune the optimized bond length, \(\mathrm{L}_{12}^{\mathrm{opt}}\), we incorporated a motorized coaxial RF phase shifter into the experimental setup. The phase 
shifter is rated for operation from DC to \(18\,\mathrm{GHz}\), with an insertion loss below \(1.0\,\mathrm{dB}\) over its rated band and an average RF power 
handling of \(100\,\mathrm{W}\). The device is normally adjusted using a manual knob; to enable automated and repeatable control, the knob was mechanically coupled 
to a stepper motor. The motor was mounted to preserve axial alignment with the adjustment mechanism and to minimize backlash during operation. The motor was driven 
using a Trinamic motion controller connected to a computer through USB, allowing the phase-shifter displacement to be controlled with custom Python scripts over 
the accessible range from \(3\,\mathrm{mm}\) to \(23\,\mathrm{mm}\).

The relationship between motor motion and phase-shifter displacement was calibrated using a high-precision digital micrometer (Asimeto IP65 digital outside micrometer). 
The micrometer spindle was placed in contact with a reference surface on the phase-shifter mechanism that translated during adjustment. The motor was then driven in 
controlled increments of microsteps, and the corresponding linear displacement was recorded. When the phase shifter was moved in a single direction, the measured 
displacement depended linearly on the number of motor microsteps. The calibration yielded

\[
\mathrm{d}=\frac{\mathrm{N}_{\mathrm{steps}}}{464}\,\mathrm{mm},
\]
where \(\mathrm{N}_{\mathrm{steps}}\) is the number of motor microsteps and \(\mathrm{d}\) is the resulting mechanical displacement. Measurements were also taken after reversing the direction of motion to characterize the effect of mechanical hysteresis and backlash. In operation, target positions were therefore approached from a consistent direction so that the calibrated single-direction linear relation could be used reproducibly. Multiple calibration trials confirmed that the displacement uncertainty was within the specified accuracy of the micrometer.
\subsection*{Hybrid CMT-Network modeling}
We consider a network of coax cables connected at $n=1,\cdots,V$ vertices (T-junction). The $V-$vertex is substituted by the nonlinear resonator (see above). 
We attach $\alpha=1,\cdots,M$ TLs to a set of these vertices that differ from the nonlinear one. The voltage (field) at each coaxial cable of length $L_B$ satisfies the one-dimensional telegraph equation,
\begin{equation}
    \left(\frac{d^2}{dx^2}+k^2\right)\psi_B(x)=0
    \label{eq:telegraph}
\end{equation}
where the position $x$ on the bond $B=(n,m)$ is defined to be $x=0 (L_B)$ on vertex $n (m)$. The solution of Eq. (\ref{eq:telegraph}) is
$\psi_{B}(x)=\phi_n\,\frac{\sin\!\big(k(L_B-x)\big)}{\sin(kL_B)}+\phi_m\,\frac{\sin(kx)}{\sin(kL_B)}$ where $\phi_n (\phi_m)$ are the 
field amplitudes at $x=0 (x=L_B)$. The field amplitude at the end point of the coax cable that is coupled to the nonlinear resonator at 
the vertex $n=V$ can be expressed in terms of the magnetic field amplitude of the resonator mode $a$ and the capacitive coupling constant 
$\gamma_r (r=1,2,3)$ as $\phi_V^{(r)}=\gamma_r a$. The field $a$ satisfies the TCMT, see Eqs. (\ref{oneres},\ref{Eq:CMT-2}). 

At the TL the field can be written as $\psi_\alpha(x_\alpha)=I_\alpha e^{-ikx_\alpha}+O_\alpha e^{ikx_\alpha}$ where $x_\alpha= 0$ at the 
vertex where the lead is attached, $I_\alpha$ ($O_\alpha$) is the amplitude of the incoming (outgoing ) wave. 
Imposing field continuity and current conservation at all vertices $n$, and taking into account Eq. (\ref{Eq:CMT-2}), we come up with 
Eqs. (\ref{GSE},\ref{MGE}) of the main text.


\bibliography{adjoint_invivo}{}

@article{Momeni2023,
  author  = {Momeni, Ali and Rahmani, Babak and Mall{\'e}jac, Matthieu and del Hougne, Philipp and Fleury, Romain},
  title   = {Backpropagation-Free Training of Deep Physical Neural Networks},
  journal = {Science},
  volume  = {382},
  number  = {6676},
  pages   = {1297--1303},
  year    = {2023},
  doi     = {10.1126/science.adi8474}
}

@article{WSWSMTAM2023,
  author  = {Wang, Tianyu and Sohoni, Mandar M. and Wright, Logan G. and Stein, Martin M. and Ma, Shi-Yuan and Onodera, Tatsuhiro and Anderson, Maxwell G. and McMahon, Peter L.},
  title   = {Image Sensing with Multilayer Nonlinear Optical Neural Networks},
  journal = {Nature Photonics},
  volume  = {17},
  pages   = {408--415},
  year    = {2023},
  doi     = {10.1038/s41566-023-01170-8}
}

@article{Papp2021,
  author  = {Papp, {\'A}d{\'a}m and Porod, Wolfgang and Csaba, Gy{\"o}rgy},
  title   = {Nanoscale Neural Network Using Non-Linear Spin-Wave Interference},
  journal = {Nature Communications},
  volume  = {12},
  pages   = {6422},
  year    = {2021},
  doi     = {10.1038/s41467-021-26711-z}
}

@article{GWLK2025,
  author  = {Guillamon, John and Wang, Cheng-Zhen and Lin, Zin. and Kottos, Tsampikos},
  title   = {In-situ physical adjoint computing in multiple-scattering electromagnetic environments for wave control},
  journal = {Nature Communications},
  year    ={2025},
  volume  ={16},
  page    ={1166},
    url     = {https://doi.org/10.1038/s41467-025-66385-5}
}

@article{ZSAF2021,
  author  = {Zangeneh-Nejad, Farzad and Sounas, Dimitrios L. and Al{\`u}, Andrea and Fleury, Romain},
  title   = {Analogue Computing with Metamaterials},
  journal = {Nature Reviews Materials},
  volume  = {6},
  pages   = {207--225},
  year    = {2021},
  doi     = {10.1038/s41578-020-00243-2}
}

@article{Solli2015,
  author  = {Solli, Daniel R. and Jalali, Bahram},
  title   = {Analog Optical Computing},
  journal = {Nature Photonics},
  volume  = {9},
  pages   = {704--706},
  year    = {2015},
  doi     = {10.1038/nphoton.2015.208}
}

@article{SMCGAE2014,
  author  = {Silva, Alexandre and Monticone, Francesco and Castaldi, Giuseppe and Galdi, Vincenzo and Al{\`u}, Andrea and Engheta, Nader},
  title   = {Performing Mathematical Operations with Metamaterials},
  journal = {Science},
  volume  = {343},
  number  = {6167},
  pages   = {160--163},
  year    = {2014},
  doi     = {10.1126/science.1242818}
}

@article{Fleury2025,
  author  = {Momeni, Ali and Rahmani, Babak and Sceller, Benjamin and Wright, Logan, G. and others},
  title   = {Training of physical neural networks},
  journal = {Nature},
  year    ={2025},
  volume  ={645},
  page    ={53},
    url     = {https://doi.org/10.1038/s41586-025-09384-2}
}

@article{LTW2009,
  author  = {Lagendijk, Ad and van Tiggelen, Bart and Wiersma, Diederik S.},
  title   = {Fifty years of Anderson localization},
  journal = {Physics Today},
  volume  = {62},
  number  = {8},
  pages   = {24--29},
  year    = {2009},
  doi     = {10.1063/1.3206091},
  url     = {https://doi.org/10.1063/1.3206091}
}

@article{SGL25,
  author  = {Spall, James and Guo, Xianxin and Lvovsky, Alexander I.},
  title   = {Training neural networks with end-to-end optical backpropagation},
  journal = {Advanced Photonics},
  year    ={2025},
  volume  ={7},
  page    ={016004},
    url     = {https://doi.org/10.1117/1.AP.7.1.016004}
}

@article{WGTKK24,
  title = {Nonlinearity-induced scattering zero degeneracies for spectral management of coherent perfect absorption in complex systems},
  author = {Wang, Cheng-Zhen and Guillamon, John and Tuxbury, William and Kuhl, Ulrich and Kottos, Tsampikos},
  journal = {Phys. Rev. Appl.},
  volume = {22},
  issue = {6},
  pages = {064093},
  numpages = {18},
  year = {2024},
  month = {Dec},
  publisher = {American Physical Society},
  doi = {10.1103/PhysRevApplied.22.064093},
  url = {https://link.aps.org/doi/10.1103/PhysRevApplied.22.064093}
}

@article{WKKK23,
  title = {Loss-Induced Violation of the Fundamental Transmittance-Asymmetry Bound in Nonlinear Complex Wave Systems},
  author = {Wang, Cheng-Zhen and Kononchuk, Rodion and Kuhl, Ulrich and Kottos, Tsampikos},
  journal = {Phys. Rev. Lett.},
  volume = {131},
  issue = {12},
  pages = {123801},
  numpages = {7},
  year = {2023},
  month = {Sep},
  publisher = {American Physical Society},
  doi = {10.1103/PhysRevLett.131.123801},
  url = {https://link.aps.org/doi/10.1103/PhysRevLett.131.123801}
}

@article{CMR2022,
  title={Shaping the propagation of light in complex media},
  author={Cao, Hui and Mosk, Allard Pieter and Rotter, Stefan},
  journal={Nat. Phys.},
  volume={18},
  number={9},
  pages={994--1007},
  year={2022},
  publisher={Nature Publishing Group UK London}
}

@article{gigan2022,
  title={Roadmap on wavefront shaping and deep imaging in complex media},
  author={Gigan, Sylvain and Katz, Ori and De Aguiar, Hilton and others},
  journal={Journal of Physics: Photonics},
  volume={4},
  number={4},
  pages={042501},
  year={2022},
  publisher={IOP Publishing}
}

@article{Fink1,
author = {Geoffroy Lerosey  and Julien de Rosny  and Arnaud Tourin  and Mathias Fink },
title = {Focusing Beyond the Diffraction Limit with Far-Field Time Reversal},
journal = {Science},
volume = {315},
number = {5815},
pages = {1120-1122},
year = {2007},
doi = {10.1126/science.1134824},
URL = {https://www.science.org/doi/abs/10.1126/science.1134824}
}

@article{Fink2,
author = {Allard P. Mosk  and Ad Lagendijk and Geoffroy Lerosey and Mathias Fink},
title = {Controlling waves in space and time for imaging and focusing in complex media},
journal = {Nature Photonics},
volume = {6},
pages = {283-292},
year = {2012},
URL = {https://doi.org/10.1038/nphoton.2012.88}
}

@article{SLM1,
  title = {Measuring the Transmission Matrix in Optics: An Approach to the Study and Control of Light Propagation in Disordered Media},
  author = {Popoff, S. M. and Lerosey, G. and Carminati, R. and Fink, M. and Boccara, A. C. and Gigan, S.},
  journal = {Phys. Rev. Lett.},
  volume = {104},
  issue = {10},
  pages = {100601},
  numpages = {4},
  year = {2010},
  month = {Mar},
  publisher = {American Physical Society},
  doi = {10.1103/PhysRevLett.104.100601},
  url = {https://link.aps.org/doi/10.1103/PhysRevLett.104.100601}
}

@article{GHRSD2024,
  title = {Detecting and Focusing on a Nonlinear Target in a Complex Medium},
  author = {Go\"{\i}coechea, Antton and H\"upfl, Jakob and Rotter, Stefan and Sarrazin, Fran\ifmmode \mbox{\c{c}}\else \c{c}\fi{}ois and Davy, Matthieu},
  journal = {Phys. Rev. Lett.},
  volume = {134},
  issue = {18},
  pages = {183802},
  numpages = {7},
  year = {2025},
  month = {May},
  publisher = {American Physical Society},
  doi = {10.1103/PhysRevLett.134.183802},
  url = {https://link.aps.org/doi/10.1103/PhysRevLett.134.183802}
}

@article{jiang2024,
  title={Coherent control of chaotic optical microcavity with reflectionless scattering modes},
  author={Jiang, Xuefeng and Yin, Shixiong and Li, Huanan and Quan, Jiamin and Goh, Heedong and Cotrufo, Michele and Kullig, Julius and Wiersig, Jan and Al{\`u}, Andrea},
  journal={Nat. Phys.},
  volume={20},
  number={1},
  pages={109--115},
  year={2024},
  publisher={Nature Publishing Group UK London}
}

@article{WGKDRGK26,
  author  = {Wang, Cheng-Zhen and Guillamon, John and Kuhl, Ulrich and Davy, Matthieu and Reisner, Mattis and
Goetschy, Arthur and Kottos, Tsampikos},
  title   = {Guiding waves through chaos: Universal bounds for targeted mode transport},
  journal = {Science Advances},
  year    ={2026},
  volume  ={12},
  page    ={eaeb1158}
}

@article{GBWL21,
  author  = {Guo, Xianxin and Barrett, Thomas D. and Wang Zhiming M. and Lvovsky, Alexander I.},
  title   = {Backpropagation through nonlinear units for the all-optical training of neural networks},
  journal = {Photonics Research},
  year    ={2021},
  volume  ={9},
  page    ={B71}
}

@article{KSBS11,
  author  = {Katz, Ori and Small, Eran and Bromberg, Yaron and Silberberg, Yaron},
  title   = {Focusing and compression of ultrashort pulses through scattering media},
  journal = {Nature Photonics},
  year    = {2011},
  volume  = {5},
  number  = {6},
  pages   = {372--377},
  doi     = {10.1038/nphoton.2011.72},
  url     = {https://www.nature.com/articles/nphoton.2011.72}
}

@article{YDLWS25,
  author  = {Yang, Han Qing and Dai, Jun Yan and Li, Hui Dong and Wu, Lijie and Shen, Zi Hang and Zhou, Qun Yan and Zhang, Meng Zhen and Wang, Si Ran and Wang, Zheng Xing and Wu, Jun Wei and Jin, Shi and Tang, Wankai and Cheng, Qiang and Cui, Tie Jun},
  title   = {Adaptively programmable metasurface for intelligent wireless communications in complex environments},
  journal = {Nature Communications},
  year    = {2025},
  volume  = {16},
  number  = {1},
  pages   = {6070},
  doi     = {10.1038/s41467-025-61409-6},
  url     = {https://www.nature.com/articles/s41467-025-61409-6}
}

@article{Galiffi2026,
  title = {Optical coherent perfect absorption and amplification in a time-varying medium},
  author = {Galiffi, E. and Harwood, A.C. and Vezzoli, S. and et al.},
  journal = {Nat. Photon.},
  volume = {20},
  pages = {163-169},
  year = {2026},
  url = {https://doi.org/10.1038/s41566-025-01833-8}
}

@article{kottos2000,
  title = {Chaotic Scattering on Graphs},
  author = {Kottos, Tsampikos and Smilansky, Uzy},
  journal = {Phys. Rev. Lett.},
  volume = {85},
  issue = {5},
  pages = {968--971},
  numpages = {0},
  year = {2000},
  month = {Jul},
  publisher = {American Physical Society},
  doi = {10.1103/PhysRevLett.85.968},
  url = {https://link.aps.org/doi/10.1103/PhysRevLett.85.968}
}

@article{aluCPA,
      title={Coherent perfect absorbers: linear control of light with light}, 
      author={Baranov, D. and Krasnok, A. and Shegai, T. et al.},
       journal = {Nat Rev Mater},
  volume = {2},
  issue = {12},
  pages = {17064},
  year = {2017},
  doi = {10.1038/natrevmats.2017.64}
}

@article{WCGNSC2011,
author = {Wenjie Wan  and Yidong Chong  and Li Ge  and Heeso Noh  and A. Douglas Stone  and Hui Cao },
title = {Time-Reversed Lasing and Interferometric Control of Absorption},
journal = {Science},
volume = {331},
number = {6019},
pages = {889-892},
year = {2011},
doi = {10.1126/science.1200735},
URL = {https://www.science.org/doi/abs/10.1126/science.1200735}
}

@article{RIS7,
  title = {Spatiotemporal Wave Front Shaping in a Microwave Cavity},
  author = {del Hougne, Philipp and Lemoult, Fabrice and Fink, Mathias and Lerosey, Geoffroy},
  journal = {Phys. Rev. Lett.},
  volume = {117},
  issue = {13},
  pages = {134302},
  numpages = {6},
  year = {2016},
  month = {Sep},
  publisher = {American Physical Society},
  doi = {10.1103/PhysRevLett.117.134302},
  url = {https://link.aps.org/doi/10.1103/PhysRevLett.117.134302}
}

@article{cizmar2015,
year = {2015},
publisher = {Nature},
volume = {9},
pages = {529},
author = {Pl\"oschner, M and Tyc, T and Cizm\'ar, T},
title = {Seeing through chaos in multimode fibres},
journal = {Nature Photonics}
}

@article{BGH2022,
   author = {Nicholas Bender and Arthur Goetschy and Chia Wei Hsu and others},
   issn = {10916490},
   issue = {41},
   journal = {Proc. Natl. Acad. Sci. U.S.A.},
   title = {Coherent enhancement of optical remission in diffusive media},
   volume = {119},
   year = {2022}
}

@Article{BYGYHC2022,
author={Bender, Nicholas and Yamilov, Alexey and Goetschy, Arthur and Y{\i}lmaz, Hasan and Hsu, Chia Wei and Cao, Hui},
title={Depth-targeted energy delivery deep inside scattering media},
journal={Nature Physics},
year={2022},
month={Mar},
day={01},
volume={18},
number={3},
pages={309-315},
issn={1745-2481},
}

@article{CKA2020,
  title={Perfect absorption in complex scattering systems with or without hidden symmetries},
  author={Chen, Lei and Kottos, Tsampikos and Anlage, Steven M},
  journal={Nat. Comm.},
  volume={11},
  number={1},
  pages={5826},
  year={2020},
  publisher={Nature Publishing Group UK London}
}

@article{FSK2017,
year = {2017},
month = {jun},
publisher = {IOP Publishing},
volume = {50},
number = {30},
pages = {30LT01},
author = {Fyodorov, Yan V and Suwunnarat, Suwun and Kottos, Tsampikos},
title = {Distribution of zeros of the S-matrix of chaotic cavities with localized losses and coherent perfect absorption: non-perturbative results},
journal = {Journal of Physics A: Mathematical and Theoretical},
}

@article{MGBYHYC2024,
  title={Delivering broadband light deep inside diffusive media},
  author={McIntosh, Rohin and Goetschy, Arthur and Bender, Nicholas and Yamilov, Alexey and Hsu, Chia Wei and Y{\i}lmaz, Hasan and Cao, Hui},
  journal={Nat. Photon.},
  pages={1--7},
  year={2024},
  publisher={Nature Publishing Group UK London}
}

@article{hsu2017correlation,
  title={Correlation-enhanced control of wave focusing in disordered media},
  author={Hsu, Chia Wei and Liew, Seng Fatt and Goetschy, Arthur and Cao, Hui and Douglas Stone, A},
  journal={Nat. Phys.},
  volume={13},
  number={5},
  pages={497--502},
  year={2017},
  publisher={Nature Publishing Group UK London}
}

@article{kottos2001,
  title={Quantum graphs: a model for quantum chaos},
  author={Kottos, Tsampikos and Schanz, Holger},
  journal={Physica E},
  volume={9},
  number={3},
  pages={523--530},
  year={2001},
  publisher={Elsevier}
}

@article{kuchment2004,
  title={Quantum graphs and their applications},
  author={Kuchment, P},
  journal={Special issue of Waves in Random Media},
  volume={14},
  number={1},
  year={2004}
}

@article{hughes2018,
  title={Training of photonic neural networks through in situ backpropagation and gradient measurement},
  author={Hughes, Tyler W and Minkov, Momchil and Shi, Yu and Fan, Shanhui},
  journal={Optica},
  volume={5},
  number={7},
  pages={864--871},
  year={2018},
  publisher={Optica Publishing Group}
}

@article{wetzstein2020,
  title={Inference in artificial intelligence with deep optics and photonics},
  author={Wetzstein, Gordon and Ozcan, Aydogan and Gigan, Sylvain and others},
  journal={Nature},
  volume={588},
  number={7836},
  pages={39--47},
  year={2020},
  publisher={Nature Publishing Group UK London}
}

@article{shen2017deep,
  title={Deep learning with coherent nanophotonic circuits},
  author={Shen, Yichen and Harris, Nicholas C and Skirlo, Scott and others},
  journal={Nature photonics},
  volume={11},
  number={7},
  pages={441--446},
  year={2017},
  publisher={Nature Publishing Group}
}

@article{wright2022,
  title={Deep physical neural networks trained with backpropagation},
  author={Wright, Logan G and Onodera, Tatsuhiro and Stein, Martin M and others},
  journal={Nature},
  volume={601},
  number={7894},
  pages={549--555},
  year={2022},
  publisher={Nature Publishing Group UK London}
}

@incollection{SM2003,
  author    = {Skipetrov, S. E. and Maynard, R.},
  title     = {Diffuse Waves in Nonlinear Disordered Media},
  editor    = {van Tiggelen, B. A. and Skipetrov, S. E.},
  booktitle = {Wave Scattering in Complex Media: From Theory to Applications},
  series    = {NATO Science Series II: Mathematics, Physics and Chemistry},
  volume    = {107},
  pages     = {75--98},
  year      = {2003},
  publisher = {Springer},
  address   = {Dordrecht},
  doi       = {10.1007/978-94-010-0227-1_5}
}

@article{SFKK2023,
  author  = {Suntharalingam, Arunn and Fern{\'a}ndez-Alc{\'a}zar, Lucas and Kononchuk, Rodion and Kottos, Tsampikos},
  title   = {Noise resilient exceptional-point voltmeters enabled by oscillation quenching phenomena},
  journal = {Nature Communications},
  volume  = {14},
  pages   = {5515},
  year    = {2023},
  doi     = {10.1038/s41467-023-41189-7},
  url     = {https://doi.org/10.1038/s41467-023-41189-7}
}

@article{HMWF18,
author = {Hughes, Tyler
W. and Minkov, Momchil and Williamson, Ian A. D. and Fan, Shanhui},
title = {Adjoint Method and Inverse Design for Nonlinear Nanophotonic Devices},
journal = {ACS Photonics},
volume = {5},
number = {12},
pages = {4781-4787},
year = {2018},
doi = {10.1021/acsphotonics.8b01522},
URL = {https://doi.org/10.1021/acsphotonics.8b01522},
eprint = {https://doi.org/10.1021/acsphotonics.8b01522}
}

@article{PSHPBWMMA2023,
  author  = {Pai, Sunil and Sun, Zhanghao and Hughes, Tyler W. and Park, Taewon and Bartlett, Benjamin and Williamson, Ian A. D. and Minkov, Momchil and Milanizadeh, Milad and Abebe, Netsanet and Morichetti, Francesco and Melloni, Andrea and Fan, Shanhui and Solgaard, Olav and Miller, David A. B.},
  title   = {Experimentally realized in situ backpropagation for deep learning in photonic neural networks},
  journal = {Science},
  volume  = {380},
  number  = {6643},
  pages   = {398--404},
  year    = {2023},
  doi     = {10.1126/science.ade8450},
  url     = {https://www.science.org/doi/10.1126/science.ade8450}
}

@article{LM23,
  title = {Self-Learning Machines Based on Hamiltonian Echo Backpropagation},
  author = {L\'opez-Pastor, V\'{\i}ctor and Marquardt, Florian},
  journal = {Phys. Rev. X},
  volume = {13},
  issue = {3},
  pages = {031020},
  numpages = {34},
  year = {2023},
  month = {Aug},
  publisher = {American Physical Society},
  doi = {10.1103/PhysRevX.13.031020},
  url = {https://link.aps.org/doi/10.1103/PhysRevX.13.031020}
}

@article{DMW2025,
  author        = {Dal Cin, Nicola and Marquardt, Florian and Wanjura, Clara C.},
  title         = {Training nonlinear optical neural networks with Scattering Backpropagation},
  journal       = {arXiv preprint arXiv:2508.11750},
  year          = {2025},
  eprint        = {2508.11750},
  archivePrefix = {arXiv},
  primaryClass  = {physics.optics},
  doi           = {10.48550/arXiv.2508.11750},
  url           = {https://arxiv.org/abs/2508.11750}
}

@article{Chen2024,
  author  = {Chen, Dong-Yan and Dong, Lei and Huang, Qing-An},
  title   = {Inductor-capacitor passive wireless sensors using nonlinear parity-time symmetric configurations},
  journal = {Nature Communications},
  volume  = {15},
  pages   = {9312},
  year    = {2024},
  doi     = {10.1038/s41467-024-53655-x}
}

@article{SFWRKK2025,
  author  = {Suntharalingam, Arunn and Fern{\'a}ndez-Alc{\'a}zar, Lucas J. and Wagner-Boi{\'a}n, Pablo Fabi{\'a}n and Reisner, Mattis and Kuhl, Ulrich and Kottos, Tsampikos},
  title   = {Symmetry-violation-driven hysteresis loops as measurands for noise-resilient sensors},
  journal = {Physical Review Applied},
  volume  = {23},
  pages   = {064043},
  year    = {2025},
  doi     = {10.1103/zm1g-xnn5}
}

@Inbook{Baum1986,
author={Baum, Carl E.},
editor={Thompson, James E. and Luessen, Lawrence H.},
title={Electromagnetic Topology for the Analysis and Design of Complex Electromagnetic Systems},
bookTitle={Fast Electrical and Optical Measurements: Volume I - Current and Voltage Measurements / Volume II - Optical Measurements},
year={1986},
publisher={Springer Netherlands},
address={Dordrecht},
pages={467--547},
isbn={978-94-017-0445-8},
doi={10.1007/978-94-017-0445-8-17},
url={https://doi.org/10.1007/978-94-017-0445-8-17}
}
\bibliographystyle{sciencemag}

\newpage

\renewcommand{\thefigure}{S\arabic{figure}}
\renewcommand{\thetable}{S\arabic{table}}
\renewcommand{\theequation}{S\arabic{equation}}
\renewcommand{\thepage}{S\arabic{page}}
\setcounter{figure}{0}
\setcounter{table}{0}
\setcounter{equation}{0}
\setcounter{page}{1} 

\clearpage

\begin{center}
\section*{Supplementary Materials for\\ \scititle}

John Guillamon$^{1,\dagger}$, 
William Tuxbury$^{1,\dagger}$, 
Cheng-Zhen Wang$^{1,\dagger}$, 
Owen Miller$^2$, 
Zin Lin$^3$, 
Tsampikos Kottos$^{1,\ast}$\and

\small$^\dagger$These authors contributed equally to this work

\small$^\ast$ Corresponding author. Email: tkottos@wesleyan.edu.

\end{center}

\subsubsection*{This PDF file includes:}
Materials and Methods\\
Figure S1 

\newpage
\renewcommand{\thesubsection}{S\arabic{subsection}}
\renewcommand{\thesubsubsection}{S\arabic{subsection}.\Alph{subsubsection}}

\subsection{Hybrid CMT--network modeling}

The nonlinear frequency-domain system used for the in-silico simulations contains three distinct components. The first is a microwave graph network made of one-dimensional coaxial bonds joining ideal tee junctions. The second is a set of semi-infinite transmission lines (TLs) attached to selected vertices and used to inject and collect radiation. The third is a single nonlinear resonant degree of freedom, described by one complex amplitude $a\in\mathbb{C}$, which couples to the graph through a finite set of internal coaxial attachment bonds. No particular nonlinear law will be assumed at first. Instead, the nonlinear resonator will be represented by a general scalar response function of the resonator intensity $|a|^2$.

The propagation constant in every coaxial segment is written as
\begin{equation}
    k = \frac{\omega n_r}{c},
    \label{eq:k_def}
\end{equation}
where $c$ is the vacuum speed of light and $n_r$ is the complex effective refractive index of the coaxial cable; the imaginary part of $n_r$ characterizes distributed Ohmic losses along the cables. The reduced system contains $V$ nodal degrees of freedom. The first $V-1$ are the ordinary graph-vertex voltages $\phi_n$, $n=1,\ldots,V-1$, and the final degree of freedom is the nonlinear resonator amplitude $a$. Thus, consistently with the notation of the main text, the full steady-state field vector is
\begin{equation}
    \mathbf{\Phi}=(\phi_1,\ldots,\phi_{V-1},a)^T,
    \qquad
    \boldsymbol{\phi}:=(\phi_1,\ldots,\phi_{V-1})^T.
    \label{eq:Phi_def_supp}
\end{equation}
Here $\boldsymbol{\phi}$ denotes only the ordinary graph part of the state, while $\mathbf{\Phi}$ always denotes the full $V$-component state used in the main text.

The starting point on every finite coaxial bond is the one-dimensional telegrapher equation in the monochromatic regime,
\begin{equation}
    \left(\frac{d^2}{dx_{nm}^2}+k^2\right)\psi_{nm}(x_{nm})=0.
    \label{eq:telegraph_supp}
\end{equation}
If a bond of length $L_{nm}$ connects two ordinary graph vertices whose reduced amplitudes are $\phi_n$ at $x_{nm}=0$ and $\phi_m$ at $x_{nm}=L_{nm}$, then the unique solution of Eq.~\eqref{eq:telegraph_supp} is
\begin{equation}
    \psi_{nm}(x_{nm})=
    \phi_n\frac{\sin\!\big(k(L_{nm}-x_{nm})\big)}{\sin(kL_{nm})}
    +
    \phi_m\frac{\sin(kx_{nm})}{\sin(kL_{nm})}.
    \label{eq:bond_general_solution_supp}
\end{equation}
Differentiating at the left endpoint gives
\begin{equation}
    \frac{1}{k}\,\psi_{nm}'(0)
    =-\phi_n\cot(kL_{nm})+\phi_m\csc(kL_{nm}),
    \label{eq:left_derivative_general_supp}
\end{equation}
while differentiating at the right endpoint and taking the derivative outward from that endpoint gives
\begin{equation}
    -\frac{1}{k}\,\psi_{nm}'(L_{nm})
    =\phi_n\csc(kL_{nm})-\phi_m\cot(kL_{nm}).
    \label{eq:right_derivative_general_supp}
\end{equation}

Let $\mathcal A$ be the symmetric adjacency matrix for the ordinary linear subnetwork, with $\mathcal A_{nm}=1$ when an ordinary coaxial bond joins vertices $n$ and $m$, and $\mathcal A_{nm}=0$ otherwise. Applying Eq.~\eqref{eq:left_derivative_general_supp} to every ordinary bond incident on vertex $n$ gives the outgoing dimensionless current from that vertex into the ordinary graph,
\begin{equation}
    \sum_{m\neq n}^{V-1}\mathcal A_{nm}
    \left[-\phi_n\cot(kL_{nm})+\phi_m\csc(kL_{nm})\right].
    \label{eq:ordinary_current_sum_supp}
\end{equation}
This motivates the ordinary-graph block
\begin{equation}
(H_{\mathrm{g}})_{nm}=
\begin{cases}
-\displaystyle\sum_{\ell\neq n}^{V-1}\mathcal A_{n\ell}\cot(kL_{n\ell}), & n=m,\\[1.2ex]
\mathcal A_{nm}\csc(kL_{nm}), & n\neq m,
\end{cases}
\qquad n,m=1,\ldots,V-1.
\label{eq:Hg_def_supp}
\end{equation}
This is the graph operator obtained before the nonlinear resonator attachment bonds are included. Below, the symbol $\mathbf H_L^{(V-1)}$ will denote the corresponding linear block after all purely linear loading terms on the ordinary graph vertices have been collected. This is the same block notation used in Eq.~\eqref{GSE} of the main text.

We next attach the external TLs. Suppose that $M$ semi-infinite TLs are attached to selected nodes of the full reduced system, including possibly the nonlinear resonator node. If TL $\alpha$ is attached to node $n$, and if its coordinate $x_\alpha\geq 0$ is measured outward from the network into the TL, then the TL field is decomposed as
\begin{equation}
    \psi_\alpha(x_\alpha)=I_\alpha e^{-ikx_\alpha}+O_\alpha e^{ikx_\alpha},
    \label{eq:lead_field_supp}
\end{equation}
where $I_\alpha$ is the wave incident on the network and $O_\alpha$ is the wave leaving the network. The boundary value at the attachment point is therefore
\begin{equation}
    \Phi_n=I_\alpha+O_\alpha,
    \label{eq:lead_boundary_supp}
\end{equation}
where $\Phi_n$ denotes the $n$th component of the full vector $\mathbf\Phi$. The corresponding outgoing dimensionless current into the TL is
\begin{equation}
    \frac{1}{k}\,\partial_{x_\alpha}\psi_\alpha(0)=i(O_\alpha-I_\alpha)=i\Phi_n-2iI_\alpha.
    \label{eq:lead_current_supp}
\end{equation}
Let $\mathbf W$ denote the $M\times V$ TL-incidence matrix, with $W_{\alpha n}=1$ when TL $\alpha$ is attached to node $n$ and $W_{\alpha n}=0$ otherwise, and collect the incident TL amplitudes into
\begin{equation}
    \mathbf I=(I_1,\ldots,I_M)^T,
    \qquad I_\alpha=A_\alpha e^{i\theta_\alpha}.
    \label{eq:I_def_supp}
\end{equation}
The TL contribution to the nodal current balance is therefore
\begin{equation}
    i\mathbf W^T\mathbf W\mathbf\Phi-2i\mathbf W^T\mathbf I.
    \label{eq:lead_matrix_term_supp}
\end{equation}
Equivalently, the source vector used in the main text is
\begin{equation}
    \mathbf b=2i\mathbf W^T\mathbf I.
    \label{eq:b_def_supp}
\end{equation}
The outgoing wave amplitudes are obtained from the nodal fields through
\begin{equation}
    \mathbf O=\mathbf W\mathbf\Phi-\mathbf I.
    \label{eq:output_relation_supp}
\end{equation}
This relation fixes the meaning of the measured reflections and transmissions once the nonlinear steady state has been found.

We now derive the contribution of the nonlinear resonator. Assume that the resonator is connected to the ordinary graph through $K$ finite coaxial attachment bonds. The $i$th attachment bond joins ordinary graph vertex $\ell_i$ to the resonator and has length $L_i$. Its coordinate $x_i$ is measured from the graph toward the resonator, so that $x_i=0$ at graph vertex $\ell_i$ and $x_i=L_i$ at the resonator. The field on this bond again satisfies Eq.~\eqref{eq:telegraph_supp}. The left endpoint value is the graph amplitude,
\begin{equation}
    \psi_i(0)=\phi_{\ell_i},
    \label{eq:attachment_left_value_supp}
\end{equation}
while the right endpoint value is fixed by the single-mode reduction of the local resonator physics. Since only one resonant degree of freedom is retained, the field launched from the resonator into attachment bond $i$ is proportional to the same scalar amplitude $a$. We therefore define the real positive coupling coefficient $\gamma_i$ by writing
\begin{equation}
    \psi_i(L_i)=\gamma_i a.
    \label{eq:attachment_right_value_supp}
\end{equation}
This is not an extra assumption about the wave equation on the bond. Rather, it defines the reduced coupling strength appearing in the single-mode hybrid model.

Equations~\eqref{eq:attachment_left_value_supp} and~\eqref{eq:attachment_right_value_supp} fix the attachment-bond field uniquely:
\begin{equation}
    \psi_i(x)=
    \phi_{\ell_i}\frac{\sin\!\big(k(L_i-x)\big)}{\sin(kL_i)}
    +
    \gamma_i a\frac{\sin(kx)}{\sin(kL_i)}.
    \label{eq:attachment_solution_supp}
\end{equation}
Differentiating at the graph end produces the outgoing dimensionless current from graph vertex $\ell_i$ into attachment bond $i$,
\begin{equation}
    \frac{1}{k}\,\psi_i'(0)
    =-\phi_{\ell_i}\cot(kL_i)+\gamma_i a\csc(kL_i),
    \label{eq:attachment_graph_current_supp}
\end{equation}
while differentiating at the resonator end and using the outward derivative from the resonator into the bond gives
\begin{equation}
    -\frac{1}{k}\,\psi_i'(L_i)
    =\phi_{\ell_i}\csc(kL_i)-\gamma_i a\cot(kL_i).
    \label{eq:attachment_res_current_supp}
\end{equation}
These two expressions determine, respectively, how the resonator attachment loads the ordinary graph and how the ordinary graph drives the resonator.

The graph-side attachment currents in Eq.~\eqref{eq:attachment_graph_current_supp} have two parts. The first part is a purely linear diagonal loading of the ordinary vertex $\ell_i$, and the second part couples the ordinary graph to the resonator amplitude $a$. It is useful to absorb the diagonal loading into the ordinary linear block by defining
\begin{equation}
    \mathbf H_L^{(V-1)}
    :=
    \mathbf H_{\mathrm g}
    -\sum_{i=1}^{K}\cot(kL_i)\,\mathbf e_{\ell_i}\mathbf e_{\ell_i}^{T},
    \label{eq:HL_block_def_supp}
\end{equation}
where $\mathbf e_n$ is the unit vector in the $n$th ordinary-graph direction. This is the ordinary-graph block appearing in the upper-left part of Eq.~\eqref{GSE}. If the attachment-bond self-loads are instead included directly in the definition of the ordinary linear subnetwork, Eq.~\eqref{eq:HL_block_def_supp} is simply understood as that collected linear block.

The remaining graph--resonator coupling terms are collected into the $(V-1)$-component vector $\mathbf G$, whose entries are
\begin{equation}
    G_n:=\sum_{i=1}^{K}\delta_{n\ell_i}\,\gamma_i\csc(kL_i),
    \qquad n=1,\ldots,V-1.
    \label{eq:G_def_supp}
\end{equation}
With these definitions, the ordinary-graph part of the nodal balance is
\begin{equation}
    \mathbf H_L^{(V-1)}\boldsymbol{\phi}+\mathbf G a
    +\left[i\mathbf W^T\mathbf W\mathbf\Phi\right]_{1:V-1}
    =\mathbf b_{1:V-1},
    \label{eq:graph_compact_supp}
\end{equation}
where the subscript $1:V-1$ denotes the first $V-1$ components of the corresponding full $V$-component vector.

It remains to derive the resonator equation and connect the above graph notation to temporal coupled-mode theory. The point of the coupled-mode reduction is that all of the local physics of the cavity--diode assembly is represented by the single complex modal amplitude $a(t)$, normalized so that $|a|^2$ is proportional to the stored modal energy. Keeping the time dependence explicit for a moment, the local resonator dynamics before eliminating the finite graph attachment bonds can be represented in the form
\begin{equation}
    \dot a=
    \left[-i\omega_r-\gamma_{\mathrm{nr}}-i f(|a|^2)\right]a
    +\text{linear drive from the attached channels},
    \label{eq:cmt_start_a_supp}
\end{equation}
where $\omega_r$ is the small-signal resonance frequency, $\gamma_{\mathrm{nr}}$ is the non-radiative linear loss rate, and $f(|a|^2)$ is a general intensity-dependent complex correction to the resonator response. Its real part gives an intensity-dependent frequency shift, while its imaginary part gives an intensity-dependent loss or gain. Passing to the monochromatic convention $a(t)=ae^{-i\omega t}$ gives a scalar frequency-domain resonator balance. In the notation of the main text, all linear terms local to the resonator are collected into $(H_L)_{VV}$, while the nonlinear contribution is kept separately as $f(|a|^2)$.

Using Eq.~\eqref{eq:attachment_res_current_supp}, the resonator-side attachment currents contribute
\begin{equation}
    \sum_{i=1}^{K}\gamma_i\left[-\frac{1}{k}\psi_i'(L_i)\right]
    =
    \sum_{i=1}^{K}\gamma_i\phi_{\ell_i}\csc(kL_i)
    -
    \sum_{i=1}^{K}\gamma_i^2\cot(kL_i)\,a.
    \label{eq:res_current_sum_supp}
\end{equation}
The first term is exactly $\mathbf G^T\boldsymbol{\phi}$, because Eq.~\eqref{eq:G_def_supp} implies
\begin{equation}
    \mathbf G^T\boldsymbol{\phi}
    =\sum_{n=1}^{V-1}G_n\phi_n
    =\sum_{i=1}^{K}\gamma_i\csc(kL_i)\phi_{\ell_i}.
    \label{eq:GT_phi_expand_supp}
\end{equation}
The second term is a linear self-loading of the resonator. Thus the linear resonator entry used in the main text may be written schematically as
\begin{equation}
    (H_L)_{VV}=h_0(\omega)-\sum_{i=1}^{K}\gamma_i^2\cot(kL_i),
    \label{eq:HL_VV_def_supp}
\end{equation}
where $h_0(\omega)$ is the intrinsic linear one-mode response of the isolated nonlinear cavity. For the experimental diode-loaded resonator used in the main text, this entry is fitted in the form
\begin{equation}
    (H_L)_{VV}=-4\pi z_0-\sum_{i=1}^{K}\gamma_i^2\cot(kL_i),
    \qquad
    f(|a|^2)=\frac{4\pi z_1}{1+\chi |a|^2}.
    \label{eq:experimental_res_response_supp}
\end{equation}
The complex parameters $z_0$, $z_1$, and $\chi$ encode the fitted linear and nonlinear response of the diode-loaded cavity.

Combining the resonator balance with the TL current contribution gives the $V$th component of the steady-state system,
\begin{equation}
    \mathbf G^T\boldsymbol{\phi}+\big[(H_L)_{VV}+f(|a|^2)\big]a
    +\left[i\mathbf W^T\mathbf W\mathbf\Phi\right]_{V}
    =b_V.
    \label{eq:resonator_compact_supp}
\end{equation}
Equations~\eqref{eq:graph_compact_supp} and~\eqref{eq:resonator_compact_supp} are exactly the component form of the compact hybrid network equation used 
in the main text,
\begin{equation}
\boxed{
\mathbf F(\mathbf\Phi,\mathbf\Phi^*,\mathbf p)=
\left(
\begin{bmatrix}
\mathbf H_L^{(V-1)} & \mathbf G\\
\mathbf G^T & (H_L)_{VV}
\end{bmatrix}
+i\mathbf W^T\mathbf W
\right)\mathbf\Phi
+f(|a|^2)\mathbf P_V\mathbf\Phi
-\mathbf b
=0.}
\label{eq:GSE_supp_expanded}
\end{equation}
Here $(P_V)_{nm}=\delta_{nV}\delta_{mV}$ is the projector onto the nonlinear resonator component. Equation~\eqref{eq:GSE_supp_expanded} also clarifies why 
the nonlinear response appears only in the $V$th component: since $\mathbf P_V\mathbf\Phi=(0,\ldots,0,a)^T$, the nonlinear term is $f(|a|^2)a$ in the resonator 
equation and zero on all ordinary graph vertices.

The rest of the derivation reduces this $V$-component nonlinear equation to a scalar self-consistency equation for the resonator intensity. To keep the 
notation transparent, partition the source vector as
\begin{equation}
    \mathbf b=\begin{bmatrix}\mathbf b_{\mathrm g}\\ b_V\end{bmatrix},
    \qquad
    \mathbf b_{\mathrm g}\in\mathbb C^{V-1}.
    \label{eq:b_partition_supp}
\end{equation}
For ordinary one-port-per-node TL attachments, $\mathbf W^T\mathbf W$ is diagonal. Then the TLs attached to ordinary graph vertices contribute only to the 
ordinary block, while TLs attached to the resonator shift only the resonator denominator. Define
\begin{equation}
    \mathbf A_{\mathrm g}(\omega)
    :=\mathbf H_L^{(V-1)}+i\left(\mathbf W^T\mathbf W\right)_{\mathrm g\mathrm g},
    \label{eq:Ag_def_supp}
\end{equation}
where $(\mathbf W^T\mathbf W)_{\mathrm g\mathrm g}$ is the upper-left $(V-1)\times(V-1)$ block, and define the scalar nonlinear resonator denominator
\begin{equation}
    h(y;\omega):=(H_L)_{VV}+i\left(\mathbf W^T\mathbf W\right)_{VV}+f(y),
    \qquad y:=|a|^2.
    \label{eq:h_y_def_supp}
\end{equation}
With these definitions, Eq.~\eqref{eq:GSE_supp_expanded} becomes the two-line system
\begin{align}
    \mathbf A_{\mathrm g}\boldsymbol{\phi}+\mathbf G a &= \mathbf b_{\mathrm g},
    \label{eq:two_line_graph_supp}\\
    \mathbf G^T\boldsymbol{\phi}+h(y;\omega)a &= b_V.
    \label{eq:two_line_res_supp}
\end{align}
If no TL is attached directly to the nonlinear resonator, then $b_V=0$ and $(\mathbf W^T\mathbf W)_{VV}=0$.

Assuming $\mathbf A_{\mathrm g}$ is invertible at the operating frequency, the first line gives
\begin{equation}
    \boldsymbol{\phi}=\mathbf A_{\mathrm g}^{-1}\mathbf b_{\mathrm g}
    -\mathbf A_{\mathrm g}^{-1}\mathbf G a.
    \label{eq:phi_in_terms_of_a_supp}
\end{equation}
Substituting this into the resonator equation gives
\begin{equation}
    \mathbf G^T\mathbf A_{\mathrm g}^{-1}\mathbf b_{\mathrm g}
    -\mathbf G^T\mathbf A_{\mathrm g}^{-1}\mathbf G\,a
    +h(y;\omega)a=b_V.
    \label{eq:res_after_substitution_supp}
\end{equation}
It is useful to define two scalar quantities computed entirely from the linear ordinary graph,
\begin{equation}
    \xi(\omega):=\mathbf G^T\mathbf A_{\mathrm g}^{-1}(\omega)\mathbf G,
    \qquad
    \eta(\omega;\mathbf I):=\mathbf G^T\mathbf A_{\mathrm g}^{-1}(\omega)\mathbf b_{\mathrm g}.
    \label{eq:xi_eta_defs_supp}
\end{equation}
The scalar resonator equation then becomes
\begin{equation}
    \big[h(y;\omega)-\xi(\omega)\big]a=b_V-\eta(\omega;\mathbf I),
    \label{eq:a_scalar_before_solution_supp}
\end{equation}
so that
\begin{equation}
\boxed{
    a=\frac{b_V-\eta(\omega;\mathbf I)}{h(|a|^2;\omega)-\xi(\omega)}.}
    \label{eq:a_explicit_eta_xi_supp}
\end{equation}
This equation shows explicitly how the ordinary graph dresses the local nonlinear resonator: the scalar $\xi$ is the graph-induced self-energy of the 
resonator, while $\eta$ is the graph-mediated drive that reaches the resonator from the external incident fields. When the resonator is not directly 
driven by an attached TL, $b_V=0$ and the numerator reduces to $-\eta$.

The ordinary graph field follows from Eq.~\eqref{eq:phi_in_terms_of_a_supp}:
\begin{equation}
\boxed{
    \boldsymbol{\phi}
    =\mathbf A_{\mathrm g}^{-1}\mathbf b_{\mathrm g}
    -\frac{b_V-\eta(\omega;\mathbf I)}{h(|a|^2;\omega)-\xi(\omega)}
    \mathbf A_{\mathrm g}^{-1}\mathbf G.}
    \label{eq:phi_explicit_in_input_supp}
\end{equation}
Together, Eqs.~\eqref{eq:a_explicit_eta_xi_supp} and~\eqref{eq:phi_explicit_in_input_supp} reconstruct the full main-text state vector $\mathbf\Phi
=(\boldsymbol{\phi},a)^T$ once the intensity $y=|a|^2$ is known.

Taking the modulus squared of Eq.~\eqref{eq:a_explicit_eta_xi_supp} gives the closed scalar self-consistency equation
\begin{equation}
\boxed{
    y =
    \frac{
        \left| b_V-\eta(\omega;\mathbf{I}) \right|^2
    }{
        \left| h(y;\omega)-\xi(\omega) \right|^2
    },
    \qquad y=|a|^2 .
}
\label{eq:y_scalar_master_supp}
\end{equation}
This is the sharpest reduction of the hybrid problem: all ordinary graph degrees of freedom have been integrated out into the two known quantities $\xi$ 
and $\eta$, while the direct resonator drive, if present, appears through $b_V$. Once a real nonnegative solution $y$ of Eq.~\eqref{eq:y_scalar_master_supp} 
has been selected, the resonator amplitude follows from Eq.~\eqref{eq:a_explicit_eta_xi_supp}, the ordinary graph field follows from 
Eq.~\eqref{eq:phi_explicit_in_input_supp}, and the measured outgoing waves follow from Eq.~\eqref{eq:output_relation_supp}.

If the nonlinear resonator response is affine in $y$, as in a Kerr-type reduced model, then
\begin{equation}
    h(y;\omega)=\beta_0(\omega)+\beta_1(\omega)y,
    \label{eq:h_affine_supp}
\end{equation}
and Eq.~\eqref{eq:y_scalar_master_supp} becomes
\begin{equation}
    y\left|\beta_0-\xi+\beta_1y\right|^2=|\zeta|^2,
    \qquad
    \zeta:=b_V-\eta.
    \label{eq:y_affine_prepoly_supp}
\end{equation}
Because $y$ is real, the modulus squared is a quadratic polynomial in $y$, and Eq.~\eqref{eq:y_affine_prepoly_supp} is equivalent to the cubic
\begin{equation}
    |\beta_1|^2y^3
    +2\,\Re\!\left[(\beta_0-\xi)\beta_1^*\right]y^2
    +|\beta_0-\xi|^2y
    -|\zeta|^2=0.
    \label{eq:y_affine_cubic_supp}
\end{equation}

For the diode-loaded resonator used in the main text, the fitted nonlinear contribution has the saturable form
\begin{equation}
    f(y)=\frac{4\pi z_1}{1+\chi y},
    \label{eq:f_saturable_main_supp}
\end{equation}
with complex fit parameters $z_1$ and, in general, $\chi$. Writing
\begin{equation}
    h(y;\omega)-\xi
    =\delta(\omega)+\frac{q}{1+\chi y},
    \qquad
    \delta(\omega):=(H_L)_{VV}+i(\mathbf W^T\mathbf W)_{VV}-\xi(\omega),
    \qquad
    q:=4\pi z_1,
    \label{eq:sat_delta_q_defs_supp}
\end{equation}
Eq.~\eqref{eq:y_scalar_master_supp} becomes
\begin{equation}
    y\left|\delta+\frac{q}{1+\chi y}\right|^2=|\zeta|^2.
    \label{eq:y_sat_prepoly_supp}
\end{equation}
Multiplying by $|1+\chi y|^2$ and defining
\begin{equation}
    U:=\delta+q,
    \qquad
    V:=\delta\chi,
    \label{eq:UV_defs_supp}
\end{equation}
we obtain
\begin{equation}
    y|U+Vy|^2-|\zeta|^2|1+\chi y|^2=0.
    \label{eq:y_sat_cubic_compact_supp}
\end{equation}
Since $y$ is real, this is again a cubic polynomial,
\begin{equation}
    |V|^2y^3
    +\left(2\Re(UV^*)-|\zeta|^2|\chi|^2\right)y^2
    +\left(|U|^2-2|\zeta|^2\Re\chi\right)y
    -|\zeta|^2=0.
    \label{eq:y_sat_cubic_supp}
\end{equation}
When $\chi$ is real, Eq.~\eqref{eq:y_sat_cubic_supp} reduces to the simpler real-$\chi$ expression obtained by expanding $(1+\chi y)^2$.

A final remark is useful. Because the nonlinear response has been kept general until the last step, Eq.~\eqref{eq:y_scalar_master_supp} may have more than one admissible real nonnegative solution $y$ at fixed drive, corresponding to multistability of the local resonator response. Such cases were not relevant for the in-silico emulation of the experimental platform, because the fitted platform did not display multistability in the operating regime used here. When multiple branches are present, the physical branch must be selected by continuity from the previous operating point or by the experimentally realized sweep history.

\subsection{Frequency-resolved characterization of the nonlinear resonator}
\begin{figure}[htbp]
  \centering
  \includegraphics[width=\linewidth]{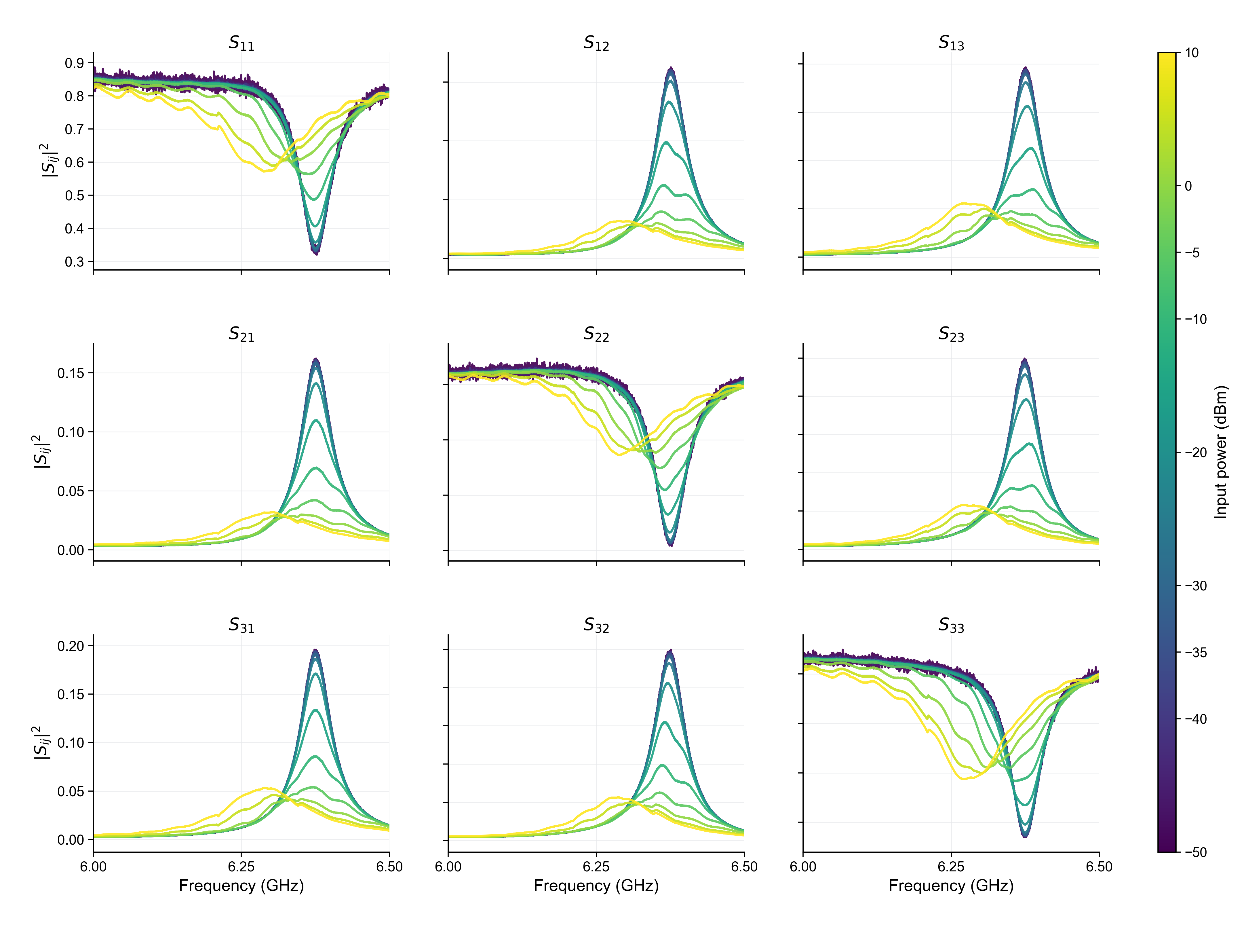}
  \caption{\textbf{Frequency-resolved scattering response of the measured multiport nonlinear vertex.}
  Scattering intensities $|S_{ij}|^2$ are plotted as a function of frequency for the measured
  input-power sweeps. Each trace corresponds to one input power from $-50$ to $10$ dBm in 5 dB
  increments, as indicated by the color scale. Rows and columns denote output and input ports,
  respectively.}
  \label{fig:supp_sparameter_frequency_slices}
\end{figure}

In Fig. \ref{fig:supp_sparameter_frequency_slices} we report the measurements of the scattering matrix elements $|S_{ij}|^2$ for $i,j=1,\cdots,3$ of the nonlinear vertex (see Fig. 2a 
of the main text) versus frequency. The various injected power levels are indicated with a color coding (see color-code bar). The vertex consists of a dielectric resonator inductively coupled with a 
ring antenna that is shorted with a diode. The coupling with the three coaxial cables occurs via three kink antennas. The whole structure is embedded inside an aluminum cylindrical resonator.

As the incident power increases, a nonlinear resonant frequency shifts towards lower frequency values is observed. At the same time, the resonant width becomes broader due to the nonlinear 
losses of the diode. The nonlinear response is also reflected in the amplitude variations of the response function as the power increases.

\end{document}